\documentclass[12pt,twoside,openright,titlepage]{report}
\usepackage{helvet}
\usepackage[T1]{fontenc}
\usepackage[latin9]{inputenc}
\usepackage[a4paper]{geometry}
\usepackage{color}
\usepackage{float}
\usepackage{amsmath}
\usepackage{amssymb}
\usepackage{graphicx}
\usepackage{setspace}
\usepackage{esint}
\usepackage{subfigure}
\makeatletter
\date{}
\usepackage{braket}
\usepackage{array}
\usepackage{minitoc}
\usepackage{nopageno}
\numberwithin{equation}{chapter}
\usepackage{fancyhdr}
    \usepackage[unicode=true,pdfusetitle,  bookmarks=true,bookmarksnumbered=false,bookmarksopen=false,  breaklinks=false,pdfborder={0 0 0},backref=false,colorlinks=true]{hyperref}

    \hypersetup{linkcolor=blue,citecolor=blue}
\newcommand{\colorin}{black}

\usepackage{cite}

\fancypagestyle{plain,empty}{%
\fancyhead{} 
}

\newcommand{\der}{\text{d}}

\makeatother

\begin{document}
\noindent \thispagestyle{plain}

\pagenumbering{roman}

\noindent %
\begin{minipage}[c]{0.3\columnwidth}%
\begin{flushleft}
    \includegraphics[scale=0.25]{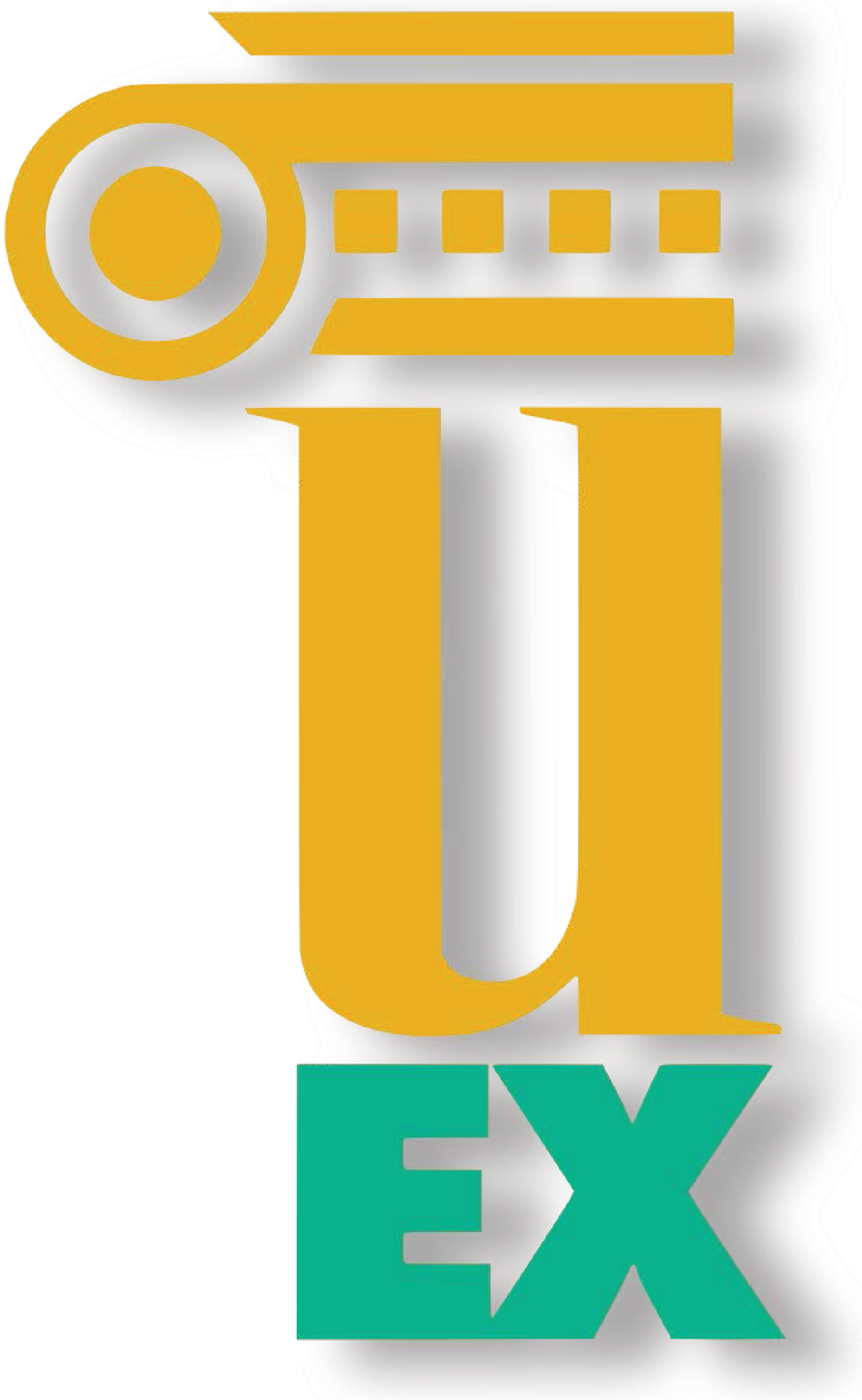}
\par\end{flushleft}%
\end{minipage}\hfill{}%
\begin{minipage}[c][0.8\textheight][t]{0.75\columnwidth}%
\noindent \begin{center}
\textbf{UNIVERSIDAD DE EXTREMADURA}\medskip{}

\par\end{center}

\noindent \begin{center}
\textbf{FACULTAD DE CIENCIAS}\medskip{}
\medskip{}
\medskip{}

\par\end{center}

\noindent \begin{center}
\textbf{Grado en F\'isica}\medskip{}
\medskip{}
\medskip{}

\par\end{center}

\noindent \begin{center}
\textbf{}%
\begin{minipage}[t]{1\columnwidth}%
\noindent \begin{center}
\textbf{MEMORIA DEL TRABAJO FIN DE GRADO}
\par\end{center}%
\end{minipage}\medskip{}
\medskip{}
\medskip{}
\medskip{}
\medskip{}
\medskip{}
\medskip{}
\medskip{}
\medskip{}

\par\end{center}

\noindent \begin{center}
\medskip{}
\medskip{}

\par\end{center}

\noindent \begin{center}
\textbf{Funciones de correlaci\'on y propiedades termof\'isicas en l\'iquidos unidimensionales}
\par\end{center}

\noindent \begin{center}
\medskip{}
\medskip{}
\medskip{}
\medskip{}
\medskip{}
\medskip{}
\medskip{}
\medskip{}
\medskip{}
\medskip{}
\medskip{}
\medskip{}
\medskip{}
\medskip{}
\medskip{}
\medskip{}

\par\end{center}

\noindent \begin{flushright}
\textbf{ANA MAR\'IA MONTERO MART\'INEZ\\SEPTIEMBRE, 2017}
\par\end{flushright}
\end{minipage}

\newpage{}

\clearpage

\thispagestyle{empty}

\mbox {}

\clearpage \newpage{}

\noindent \thispagestyle{empty}\medskip{}
\medskip{}
\medskip{}
\medskip{}
\medskip{}
\medskip{}
\medskip{}
\medskip{}
\medskip{}
\medskip{}
\medskip{}
\medskip{}
\medskip{}
\medskip{}
\medskip{}
\medskip{}

ANDR\'ES SANTOS REYES, profesor del Departamento de F\'ISICA de la Universidad
de Extremadura.

\noindent \medskip{}
\medskip{}

\noindent INFORMA:\medskip{}

\noindent \medskip{}

\noindent Que D\~na. ANA MAR\'IA MONTERO MART\'INEZ ha realizado
bajo mi direcci\'on el Trabajo Fin de Grado \textit{Funciones de correlaci\'on y propiedades termof\'isicas en l\'iquidos unidimensionales}. Considero que la memoria re\'une los
requisitos necesarios para su evaluaci\'on.\medskip{}
\medskip{}

\noindent \begin{center}
Badajoz, 1 de septiembre de 2017
\par\end{center}

\begin{center}
\medskip{}
\medskip{}
\medskip{}
\medskip{}
\medskip{}
\medskip{}
\medskip{}
\medskip{}
\medskip{}
\medskip{}
\medskip{}
\medskip{}
\medskip{}
\medskip{}
\medskip{}
\medskip{}

\par\end{center}

\noindent \begin{center}
Fdo. Andr\'es Santos Reyes
\par\end{center}

\newpage{}

\clearpage

\thispagestyle{empty}

\mbox {}

\clearpage \newpage{}

\thispagestyle{empty}






\clearpage \newpage{}

\thispagestyle{empty}

\pagenumbering{arabic}

\renewcommand{\contentsname}{Index}

\noindent \tableofcontents
\thispagestyle{plain}\newpage{}


\chapter*{Resumen}
\markboth{Abstract}{}
Se han estudiado las propiedades de los l\'iquidos unidimensionales para varios potenciales de interacci\'on para los cuales, bajo ciertas condiciones, las propiedades del sistema admiten soluciones anal\'iticas. Los potenciales estudiados son el pozo triangular y el potential rampa y, m\'as tarde, se ha utilizado el conocimiento de estos para estudiar los potenciales de esferas duras y de esferas duras adhesivas. Para cada uno de los potenciales, se ha estudiado su ecuaci\'on de estado y otras propiedades termodin\'amicas tales como el factor de compresibilidad y la energ\'ia interna. Despu\'es, se han estudiado sus funciones de correlaci\'on tales como la funci\'on de distribuci\'on radial, el factor de estructura y la funci\'on de correlaci\'on directa. Para esta \'ultima, las aproximaciones conocidas como Percus--Yevick y \textit{hypernetted-chain} han sido comparadas con el resultado anal\'itico. Finalmente, se ha estudiado el comportamiento asint\'otico de la funci\'on de distribuci\'on radial y se ha calculado la l\'inea de Fisher--Widom para el potencial pozo triangular.

{\let\clearpage\relax\chapter*{Abstract}}
We study the properties of one-dimensional liquids for several interaction potentials for which, under certain assumptions, the properties of the system admit an analytical solution. The studied potentials are the triangle-well and the ramp potential, and then we use the knowledge of these ones to study the hard-rod and the sticky-hard-rod potentials. For each one of these potentials, we study its equation of state and other thermodynamic quantities such as the compressibility factor and the internal energy. Then, we study its correlation functions, such as the radial distribution function, the structure factor and the direct correlation function. For the latter one, the approximations known as Percus--Yevick and hypernetted-chain have been compared to the analytical result. Finally, the asymptotic behaviour of the radial distribution function is studied and the Fisher--Widom line computed for the triangle-well potential.

\addcontentsline{toc}{chapter}{Abstract}
\chapter*{Objectives}
\markboth{Objectives}{}
The main objective in this project to study the theory of classical liquids and to learn how to use it in order to analyse the properties of certain one-dimensional liquids. In this sense, the study of the chosen potentials, namely the triangle-well and the ramp potentials, include the following objectives:
\begin{itemize}
	\item To obtain analytical results for certain one-dimensional potentials using statistical mechanics, such as the equation of state and the radial distribution function.
	\item To study simpler potentials like the hard-rod or the sticky-hard-rod potentials through the results obtained in the triangle-well case by taking the appropriate limits.
	\item To analyse how well the Percus--Yevick and the hypernetted-chain approximations for the direct correlation function match the analytical result in the triangle-well potential.
	\item To study the large-distance asymptotic behaviour of the radial distribution function in the triangle-well potential.
	\item To compute the Fisher--Widom line and to study its behaviour when we change the range of the triangle-well.
\end{itemize}
\addcontentsline{toc}{chapter}{Objectives}

\chapter{Introduction}

In the absence of interaction among the particles, all fluids behave like an ideal gas. However, particles in a system are always under the influence of certain interaction forces among them, which are responsible for the deviation from the ideal gas properties.

The complete description of a three-dimensional system where a certain interaction potential among its particles exists is, in general, a formidable task. To simplify the problem, several hypotheses can be made, such as pairwise additive potential but, even taking into account this simplification, three-dimensional systems cannot be solved exactly and one must use simulation methods.

In general, if we are interested in finding the analytical solution of a system of particles, we must consider only one-dimensional systems (where particles are arranged in a line) and, even in this simple situations, take at least three hypotheses to simplify the problem:
\begin{itemize}
	\item The order of the particles is fixed, which means that particles cannot exchange places in the line.
	\item The interaction must have a finite range.
	\item Each particle interacts only with its two nearest neighbours.
\end{itemize}

\begin{figure}[htbp]
	\centering
	\subfigure[Triangle well]{\includegraphics[height=4.8cm]{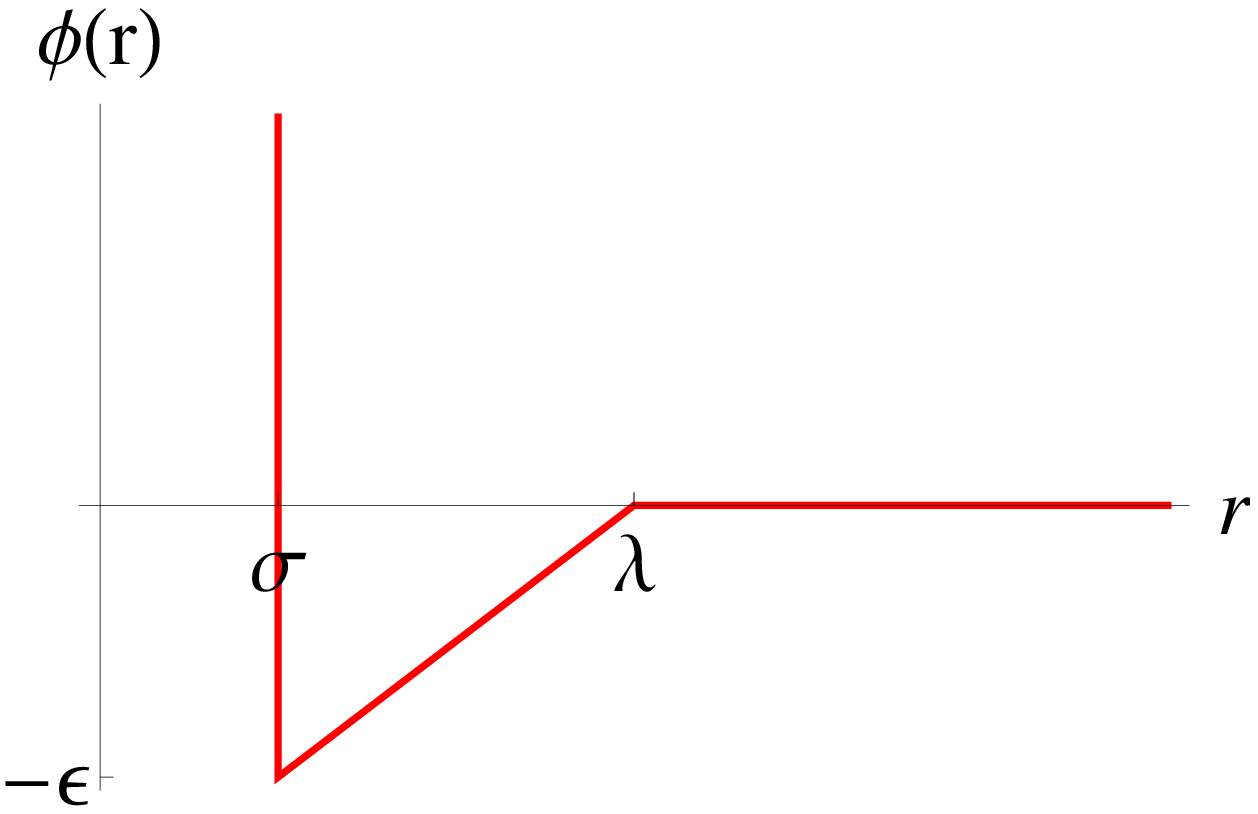} \label{trianglewell}}
	\subfigure[Ramp]{\includegraphics[height=4.8cm]{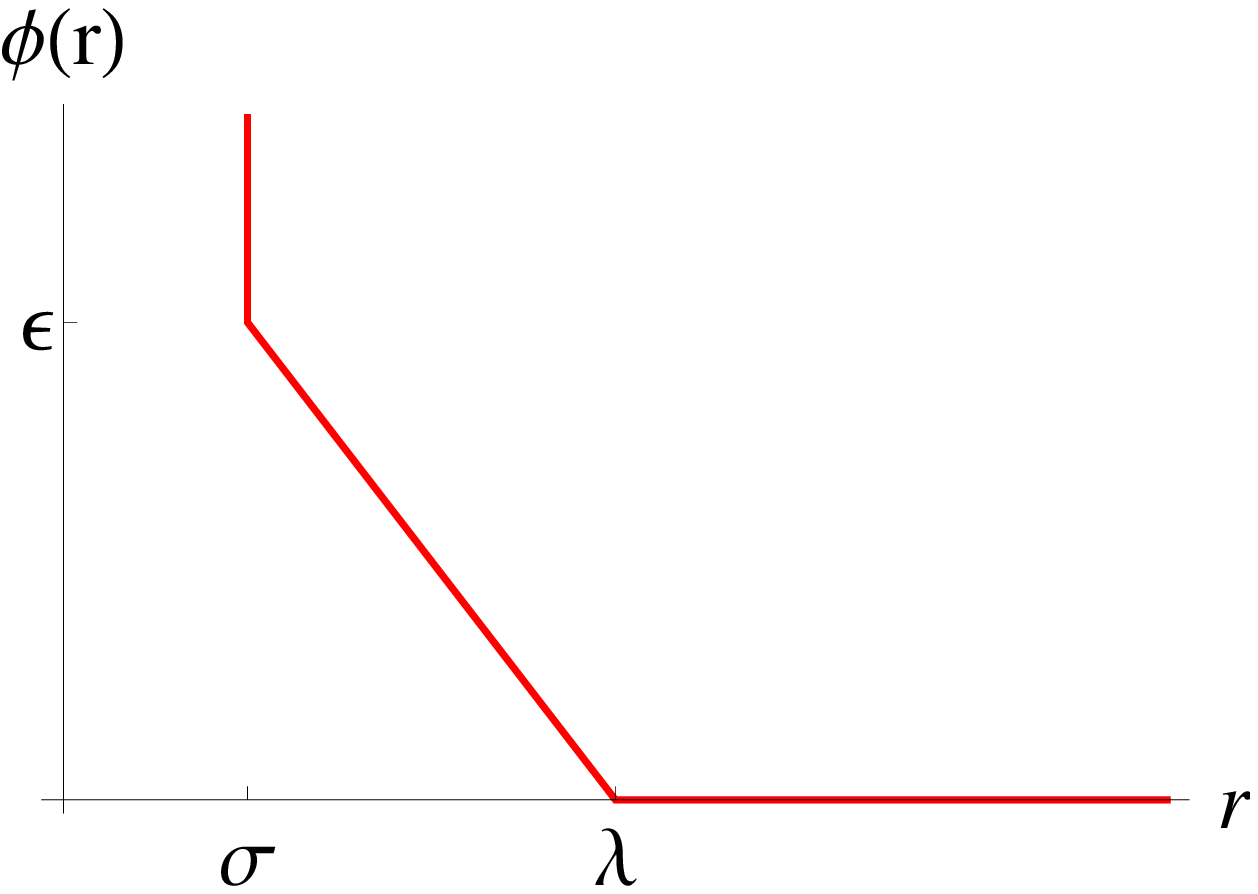} \label{ramp}}
	\caption{Plot of the two potentials to be studied in this work.}
	\label{studied_potentials}
\end{figure}

Taking into account the hypotheses above, one-dimensional systems can be exactly solved \cite{doi:10.1063/1.1699116,doi:10.1063/1.1675348,PhysRevE.76.062201,doi:10.1063/1.3256234,Hill,Andres}. Among all the possible potentials that fulfil these features, we will focus on two main potentials, the triangle-well potential and the ramp potential, both shown in  Fig.~\ref{studied_potentials}.

The triangle-well potential was already studied by Takeo Nagayima in 1940~\cite{nagayima0, nagayima} but here we will use different methods to achieve analytical results and carry out a deeper and further analysis of the liquid properties, obtaining results that were not studied back then. Furthermore, properties of other interaction potentials, such as the sticky-hard-rod and the hard-rod models, can be deduced from the triangle-well potential.

It may seem, though, as if these kinds of one-dimensional systems are far beyond what we can expect from reality and that they can only be used for didactic purposes. However, although it is true that one-dimensional systems have a great value as a learning material for students, the interest in these systems comes from applications that are far beyond the merely didactic ones:

\begin{enumerate}
	\item One-dimensional systems can be seen as three-dimensional systems confined in a very narrow tube. In this situations, one-dimensional models have proved to be a very valid approximation whose results are close to those provided by two-dimensional or three-dimensional models ~\cite{doi:10.1021/ac503880j}. Therefore, it is not surprising that one-dimensional systems could have a wide range of applications in nature such as carbon nanotubes~\cite{doi:10.1021/nn200222g,doi:10.1063/1.3593064} or biological ion-channels~\cite{BODA20083486}.
	
	\item The analytical solutions for several one-dimensional potentials serve as a reliable background from which it is possible to develop ideas that are necessary to perform the study and analysis of the three-dimensional case~\cite{PhysRev.50.955}.
	
	\item The fact that there exists an analytical solution to the one-dimensional problem provides an opportunity for a direct comparison with certain computer simulations to prove their validity~\cite{bishop11,doi:10.1063/1.1681165} and also some perturbation theories~\cite{doi:10.1063/1.450921,doi:10.1063/1.445837}.
\end{enumerate}

\vspace{10pt}

The plan of the rest of this project is as follows. The first part of the project focuses on the theoretical foundations and comprises chapters~\ref{chap1} and \ref{chap:isothermal_isobaric_ensemble}. In Chap.~\ref{chap1} we recall the main spatial correlation functions, while in Chap.~\ref{chap:isothermal_isobaric_ensemble} we develop the main statistical-mechanical concepts that are going to be needed afterwards.
The second part of the project comprises chapters~\ref{chap:tw}--\ref{chap:last} and contains the original part of the project. Chapters~\ref{chap:tw} and \ref{chap:ramp} analyse the properties of the triangle-well and ramp potentials, respectively, and Chap.~\ref{chap:last} focuses on the hard-rod and the sticky-hard-rod potentials.

\chapter{Spatial correlation functions}\label{chap1}

In the absence of interactions in a system containing $N$ particles confined in a volume $V$, every system behaves like an ideal gas one. The lack of forces among the particles leads to a disordered structure in the system, regardless of the length scale from which we are looking at it.

However, when an interaction potential exists, the fluid is forced to have a certain order. Even when a homogeneous system is considered, there is a short-range order among the closest particles due to how a particle interacts with its nearest neighbours. Those interactions usually disappear when particles are far away from each other since the interaction potential is short-range, yielding a more disordered structure as we move further away from a given reference particle.

This ordered structure in the fluid can be represented through the spatial correlation functions of the system, which are different ways of measuring the order of a system. Among those spatial correlation functions we can mention the radial distribution function, the structure factor or the direct and total correlation functions.

\section{Radial distribution function}

Let us consider a system containing $N$ particles inside a volume $V$  in the thermodynamic limit $\left(N \rightarrow \infty, \hspace{2pt} V \rightarrow \infty, \hspace{2pt} N/V=\mathrm{finite} \right)$. Its average number density of particles is then defined as $n=N/V$.

The radial distribution function in a system of particles describes how density changes with the distance measured from a reference particle located at the origin, relative to the average density.
It then is possible to define the radial distribution function $g(r)$ as~\cite{Andres, Balescu,Barrio, Helfand}
\begin{equation}\label{rdf_def}
	n_0(r)= ng(r),
\end{equation}
where $n_0(r)$ is the time-averaged local density at a distance $r$ from a certain particle located at the origin ($r=0$) when the system is isotropic and homogeneous, that is, it is rotationally and translationally invariant, respectively.

With the radial distribution function, it is possible to calculate the number of particles (among the remaining $N-1$ particles) to be found within a volume element $\text{d}\mathbf{r}$ around a point $\mathbf{r}$ as
\begin{equation}\label{dnr}
  	\der n_0(\mathbf{r}) = n_0(r) \der \mathbf{r} = n g(r) \der \mathbf{r} .
\end{equation}

\section{Total correlation function}

Given that the radial distribution function goes to 1 for very large values of $r$ (since the density around a reference particle approaches the average value for very large distances), sometimes it is convenient to use another function, $h(r)$, called the total correlation function, which measures how much two particles at a distance $r$ are correlated with each other.
It is defined through $g(r)$ as
\begin{equation*}
	h(r) = g(r) - 1 .
\end{equation*}
This definition ensures that $\lim_{r \rightarrow \infty}h(r)=0$, which can be convenient to use for some applications that require certain integrals.

\section{Structure factor}

In general, the structure factor of a system is a function that describes how a material scatters radiation. The structure factor of a fluid is interesting because it is experimentally accessible since it is directly related to the observable intensity of radiation scattered by the fluid~\cite{Balescu}. An important property of the total correlation function is that it can be related to the structure factor of the system by means of the Fourier transform.

In order to do this, let us notice that it is possible to count the number of particles in the system around a reference particle (here labelled as $i=0$) as
\begin{equation}\label{ssf1}
	\der n_0(\mathbf{r}) = \left\langle \sum_{i \neq 0} \delta(\mathbf{r}-\mathbf{r}_i) \right\rangle \der \mathbf{r} ,
\end{equation}
where the brackets mean the average in the required ensemble. The average does not depend on the specific particle $i$ chosen, so the sum over all $N-1$ particles will equal the average over one single particle multiplied by $N-1$. Because of this, and using equation~(\ref{dnr}), it is possible to write
\begin{equation}\label{ssf2}
	g(r) = \frac{1}{n} \left\langle \sum_{i \neq 0} \delta(\mathbf{r}-\mathbf{r}_i) \right\rangle = \frac{V(N-1)}{N}\left\langle \delta(\mathbf{r}-\mathbf{r}_i) \right\rangle .
\end{equation}

Now, starting from the definition of the static structure factor, $\tilde{S}(k)$, and applying equations~(\ref{ssf1}) and~(\ref{ssf2}), it is possible to obtain a relationship between $\tilde{S}(k)$ and $g(r)$ as follows:
\begin{align}
	\tilde{S}(k)&=\frac{1}{N} \left\langle \sum_{i,j}^{N} e^{-\imath \mathbf{k} \cdot(\mathbf{r}_i-\mathbf{r}_j)} \right\rangle -n\delta(\mathbf{k}) = 1 + \frac{1}{N} \left\langle \sum_{i \neq j} e^{-\imath \mathbf{k} \cdot (\mathbf{r}_i-\mathbf{r}_j)} \right\rangle -n\delta(\mathbf{k}) \nonumber \\
	&=1 + \frac{1}{N} \left\langle \int_{V} \sum_{i \neq j} \der \mathbf{r} e^{-\imath \mathbf{k} \cdot \mathbf{r}} \delta [\mathbf{r}-(\mathbf{r}_i-\mathbf{r}_j)] \right\rangle  -n\delta(\mathbf{k}) \nonumber \\
	& =1 + \frac{N(N-1)}{N} \int_{V} \der \mathbf{r} e^{-\imath \mathbf{k} \cdot \mathbf{r}} \left\langle \delta (\mathbf{r}-\mathbf{r}_1) \right\rangle -n\delta(\mathbf{k}) \nonumber \\
	& =1 + n \int_{V} \der \mathbf{r} e^{-\imath \mathbf{k} \cdot \mathbf{r}} g(r)-n\delta(\mathbf{k}) .
\end{align}
Using the integral representation of the Dirac delta, it is possible to write the structure factor in terms of the Fourier transform of the total correlation function, $\tilde{h}(k)$, as

\begin{equation}\label{eq:Sk}
\tilde{S}(k) =1 + n \int_{V} \der \mathbf{r} e^{-\imath \mathbf{k}\cdot \mathbf{r}} [g(r)-1] = 1+ n \tilde{h}(k) .
\end{equation}

\section{Direct correlation function}

The total correlation function measures the spatial correlations between two particles $1$ and $2$ that are a distance $r$ apart. But these correlations are not only due to the direct influence between particles 1 and 2, but it also measures how particle 1 affects particle 2 through a third particle (labeled 3). It is then important to separate these two contributions into the direct and the indirect one.

The direct correlation function between two particles (1 and 2), denoted as $c(r_{12})$, was defined by L. S. Ornstein and F. Zernike as~\cite{Andres, Balescu,Barrio, Helfand}
\begin{equation}\label{OZ-equation}
	h(r_{12}) = c(r_{12}) + n \int \der \mathbf{r}_3 c(r_{13})h(r_{32}) \enskip.
\end{equation}
The idea behind this equation is simple: the total correlation function between 1 and 2 is given by the sum of the direct influence and the indirect influence through a third particle, evaluated in all possible positions of that third particle. Equation (\ref{OZ-equation}) can be rewritten using the convolution properties of the Fourier transform as
\begin{equation}\label{eq:FT-OZ}
\tilde{h}(k) = \tilde{c}(k)+n \tilde{c}(k) \tilde{h}(k),
\end{equation}
where $\tilde{c}(k)$ is the Fourier transform of $c(r)$. With (\ref{eq:Sk}) and (\ref{eq:FT-OZ}) the relationships between $\tilde{h}(k)$, $\tilde{c}(k)$ and $\tilde{S}(k)$ are easy to obtain:
\begin{equation}\label{eq:relationsCF}
	\tilde{h}(k) =\frac{\tilde{c}(k)}{1-n \tilde{c}(k)}, \quad \tilde{c}(k) =\frac{\tilde{h}(k)}{1+n \tilde{h}(k)}, \quad \tilde{S}(k) =\frac{1}{1-n \tilde{c}(k)}.
\end{equation}

\section{One-dimensional fluids with nearest-neighbour interactions}\label{sec:nearest_neighbour_interaction}

The exact evaluation of the radial distribution function for a certain system is practically impossible. However, when we take into account one-dimensional systems where interactions are reduced to only the two nearest neighbours, it is possible to evaluate analytically the expression for $g(r)$ \cite{Andres}.

Let us consider an interaction potential $\phi(r)$ that has the following properties:
\begin{itemize}
	\item The order of the particles does not change, that is, $\lim_{r \rightarrow 0}\phi (r)=\infty$.
	\item The interaction has a finite range.
	\item The interaction is reduced to the two nearest neighbours.
\end{itemize}

Under these circumstances, the total potential energy of the system can be calculated by summing the interaction potential between all consecutive particles:
\begin{equation}\label{NN_potential}
	\Phi(\mathbf{r}^N)=\sum_{i =1}^{N-1}\phi(x_{i+1}-x_{i}) .
\end{equation}

Given a certain particle, in general, it is also possible to define $p^{(l)}(r) \der r$ as the conditional probability of finding its $l$th neighbour at a distance between $r$ and $r + \der r$. The integral over the entire length of the system must be equal to $1$ since the total probability of finding the $l$th neighbour somewhere inside the volume of the system must be equal to unity. Thus,
\begin{equation}\label{Pr:normalization1}
\int_0^{\infty} \der r p^{(l)}(r)=1 .
\end{equation}
Also, the sequence of probability densities $p^{(l)}(r)$ obeys the following recurrence relation:
\begin{equation}
	p^{(l)}(r)=\int_0^r  \der r' p^{(1)}(r')p^{(l-1)}(r-r'),
\end{equation}
which, due to the convolution properties of the Laplace transform, can be written as
\begin{equation}
	\hat{P}^{(l)}(s) = \hat{P}^{(1)}(s)\hat{P}^{(l-1)}(s) = \left[ \hat{P}^{(1)}(s) \right]^l ,
\end{equation}
where $\hat{P}^{(l)}(s)$ is the Laplace transform of $p^{(l)}(r)$ defined as
\begin{equation}
	\hat{P}^{(l)}(s)= \int_0^{\infty} \der r e^{-rs} p^{(l)}(r).
\end{equation}
Another form of expressing the normalization condition~(\ref{Pr:normalization1}) that will be useful in further calculations is
\begin{equation}\label{Pr:normalization2}
	\hat{P}^{(l)}(0)=1 .
\end{equation}

Now, using the definition of $n g(r) \der r$ as the total number of particles at a distance between $r$ and $r+ \der r$ [see (\ref{dnr})], we have that
\begin{equation}
	n g(r)=\sum_{l=1}^{\infty} p^{(l)}(r).
\end{equation}
Thus, introducing the Laplace transform of $g(r)$,

\begin{align} \label{G(s):calculation}
	\hat{G}(s)=& \int_{0}^{\infty}  \der r e^{-rs}g(r)
	= \frac{1}{n} \int_{0}^{\infty} \der re^{-rs} \sum_{l=1}^{\infty} p^{(l)}(r) \nonumber \\
	= &\frac{1}{n} \sum_{l=1}^{\infty} \int_{0}^{\infty} \der r e^{-rs} p^{(l)}(r)
	= \frac{1}{n} \sum_{l=1}^{\infty} \hat{P}^{(l)}(s) \nonumber \\
	=&\frac{1}{n} \sum_{l=1}^{\infty} \left[ \hat{P}^{(1)}(s) \right]^l
	= \frac{1}{n} \frac{\hat{P}^{(1)}(s)}{1-\hat{P}^{(1)}(s)}.
\end{align}
In view of the results above, it is possible to calculate the radial distribution function if the nearest-neighbour distribution function $p^{(1)}(r)$ is known.

\chapter{Isothermal-isobaric ensemble}\label{chap:isothermal_isobaric_ensemble}

\section{Thermodynamic potentials}\label{sec3.1}
An isothermal-isobaric ensemble describes a system where the independent thermodynamic variables are the temperature, $T$, the pressure, $p$, and the number of particles, $N$. On the other hand, the volume, $V$ and the energy $E$ are fluctuating quantities (only their average values are fixed).

It is possible to define the free enthalpy (or Gibbs free energy) as the following Legendre transformation:
\begin{equation}\label{gibbs:abs}
G(T,p,N) \equiv F+ pV = \mu N,
\end{equation}
where $\mu$ is the chemical potential and $F$ is the Helmholtz free energy given by

\begin{subequations}
\begin{equation}\label{helmholtz:abs}
F(T,V,{N_{\nu}}) \equiv E - TS = -pV+ \mu N,
\end{equation}
\begin{equation}\label{helmholtz:diff}
\der F = -S \der T - p \der V +  \mu\der N,
\end{equation}
\end{subequations}
$S$ being the entropy of the system.

The differential form of the Gibbs free energy can be obtained by differentiating~(\ref{gibbs:abs}) and imposing~(\ref{helmholtz:diff}):

\begin{equation}\label{gibbs:diff}
\der G = -S \der T + V \der p + \mu \der N ,
\end{equation}
so that
\begin{equation}\label{gibbs:partials}
	S=-\left( \frac{\partial G}{\partial T} \right)_{p,N}, \quad  V=\left(\frac{\partial G}{\partial p} \right)_{T,N}, \quad   \mu=\left(\frac{\partial G}{\partial N} \right)_{T,p} =\frac{G}{N}.
\end{equation}

Equation~(\ref{gibbs:diff}) shows that the natural variables for the Gibbs free energy are temperature, pressure and the number of particles. These variables match with the ones needed in the isothermal-isobaric ensemble, which makes the Gibbs free energy the appropriate thermodynamic potential to describe the ensemble.

Note that, combining (\ref{gibbs:abs}) and (\ref{helmholtz:abs}), one has
\begin{equation}\label{Entropy_thermodynamics}
S =- \frac{G}{T}+\frac{E}{T}+\frac{pV}{T} .
\end{equation}

\section{Phase-space probability density function}\label{sec3.2}
Due to the fact that it is practically impossible to describe a system with a huge number of particles from a microscopic point of view, a statistical description is needed. We will work with a classical system containing $N$ indistinguishable particles that are enclosed inside a volume $V$, so that each microstate of the system is described by its phase-space coordinates:
\begin{itemize}
	\item $N$ position vectors $\textbf{r}^N = \{ \textbf{r}_1,\textbf{r}_2,...,\textbf{r}_N \}$.
	\item $N$ momentum vectors $\textbf{p}^N = \{ \textbf{p}_1,\textbf{p}_2,...,\textbf{p}_N \}$.
\end{itemize}

To make the notation easier, we will use the following notation for the phase-space coordinates:
\[
\textbf{x}^N = \{\textbf{r}^N, \textbf{p}^N\}, \quad \text{d}\textbf{x}^N = \text{d}\textbf{r}^N \text{d}\textbf{p}^N .
\]

The phase-space probability density $\rho_N(\textbf{x}^N)$ is defined in such a way that the probability of finding the system in a microstate which lies inside a volume $\der \textbf{x}^N$ around the point $\textbf{x}^N$ can be written as $\rho_N(\textbf{x}^N)\der\textbf{x}^N$. The basic postulate in statistical mechanics says that, out of all the different probability density functions that are consistent with the constraints in the system, the one reached at equilibrium is the one that maximizes the Gibbs entropy functional, given (in the isothermal-isobaric ensemble) by
\begin{equation}\label{Gibbs_Entropy}
\mathcal{S}[\rho_N] =-k_B \int_{0}^{\infty}\der V \int \der \textbf{x}^N\rho_N(\textbf{x}^N)\ln[C_NV_0\rho_N(\textbf{x}^N)] ,
\end{equation}
where $C_N$ and $V_0$ are constants. In order to find the probability density function that maximizes the entropy functional, Lagrange's multipliers method will be used. The constraints in our system are the following ones:
\begin{subequations}
	\begin{equation}\label{constraint1}
	\int_{0}^{\infty} \der V \int \der \textbf{x}^N \rho_N(\textbf{x}^N)=1,
	\end{equation}
	\begin{equation}\label{constraint2}
	\int_{0}^{\infty} \der V \int \der \textbf{x}^N H_N(\textbf{x}^N) \rho_N(\textbf{x}^N)=\langle E \rangle,
	\end{equation}
	\begin{equation}\label{constraint3}
	\int_{0}^{\infty} \der V V \int \der \textbf{x}^N \rho_N(\textbf{x}^N)=\langle V \rangle.
	\end{equation}
\end{subequations}
Thus, the Lagrangian of the system can be written as
\begin{align}\label{lagrangemultipliers}
\mathcal{L}[\rho_N(\textbf{x}^N)] =
& -k_B \int_{0}^{\infty}\der V \int \der \textbf{x}^N\rho_N(\textbf{x}^N)\ln[C_NV_0\rho_N(\textbf{x}^N)] \nonumber \\
& - k_B \lambda \left[ \int_{0}^{\infty} \der V \int \der \textbf{x}^N \rho_N(\textbf{x}^N)-1 \right] \nonumber \\
& - k_B \beta \left[ \int_{0}^{\infty} \der V \int \der \textbf{x}^N H_N(\textbf{x}^N) \rho_N(X^N) - \langle E \rangle \right] \nonumber \\
& - k_B \gamma \left[ \int_{0}^{\infty} \der V V \int \der \textbf{x}^N \rho_N(\textbf{x}^N)-\langle V \rangle \right],
\end{align}
where $\lambda$, $\beta$ and $\gamma$ are Lagrange multipliers. Differentiating (\ref{lagrangemultipliers}) with respect to $\rho_N(\textbf{x}^N)$, making it equal to zero and solving for $\rho_N(\textbf{x}^N)$, we obtain
\begin{align}
\frac{\delta \mathcal{L}[\rho_N]}{\delta \rho_N} =
& -k_B \int_{0}^{\infty} \der V \int \der \textbf{x}^N \ln[C_N V_0 \rho_N(\textbf{x}^N)]
-k_B \int_{0}^{\infty} \der V \int \der \textbf{x}^N  \nonumber \\
& -k_B \lambda \int_{0}^{\infty} \der V \int  \der \textbf{x}^N
- k_B \beta \int_{0}^{\infty} \der V \int \der \textbf{x}^N  H_N(\textbf{x}^N) \nonumber \\
& - k_B \gamma \int_{0}^{\infty} \der V  V \int \der \textbf{x}^N=0.
\end{align}
Since the integrand must vanish, we have
\begin{equation}
\ln[C_N V_0 \rho_N(\textbf{x}^N)] + 1 + \lambda + \beta H_N(\textbf{x}^N) + \gamma V=0
\end{equation}
and, finally
\begin{equation}\label{rho_N}
\rho_N(\textbf{x}^N)=\frac{ \exp\left[-1-\lambda-\beta H_N(\textbf{x}^N)-\gamma V \right]}{C_N V_0}.
\end{equation}

If we now substitute~(\ref{rho_N}) into~(\ref{constraint1}), we get that

\begin{align}
&\int_{0}^{\infty} \der V \int \der \textbf{x}^N \frac{ \exp\left[-1-\lambda-\beta H_N(\textbf{x}^N)-\gamma V \right]}{C_N V_0}=1 ,\\
\intertext{which yields}
& \frac{e^{-1-\lambda}}{C_N V_0} = \frac{1}{\int_{0}^{\infty} \der V e^{- \gamma V} \int  \der \textbf{x}^N e^{-\beta H_N(\textbf{x}^N)}} .
\end{align}

Finally, the phase-space probability distribution function for the isothermal-isobaric ensemble at equilibrium can be written as

\begin{equation}\label{II_probabilityfunction}
\rho^{eq}_N(\textbf{x}^N)=\frac{e^{-\beta H_N-\gamma V}}{\int_{0}^{\infty} \der V e^{- \gamma V} \int \der \textbf{x}^N e^{-\beta H_N} } \equiv \frac{e^{-\beta H_N-\gamma V}}{C_N V_0 \Delta_N(\beta, \gamma)} ,
\end{equation}
where the last equality defines the isothermal-isobaric partition function $\Delta_N(\beta,\gamma)$.

In order to calculate the equilibrium entropy in the isothermal-isobaric ensemble, we substitute the probability  distribution function, $\rho^{eq}_N(\textbf{x}^N)$ into (\ref{Gibbs_Entropy}) because it is possible to identify the value of Gibbs entropy functional once the equilibrium function is known with the equilibrium entropy as seen in thermodynamics. Thus, we obtain the following expression for the entropy at equilibrium:
\begin{align}
S
& = -k_B \int_{0}^{\infty} \der V \int  \der \textbf{x}^N \rho_N(\textbf{x}^N) \left[ -\gamma V -\beta H_N -\ln\Delta_N(\beta, \gamma) \right] \nonumber \\
& = k_B \left[ \ln\Delta_N(\beta, \gamma)  + \beta \langle E \rangle + \gamma \langle V \rangle  \right] .
\end{align}
Comparing this result with (\ref{Entropy_thermodynamics}) and identifying terms, we get
\begin{equation}
	\beta = \frac{1}{k_B T} , \quad  \gamma = \beta p \\
\end{equation}
and
\begin{equation}
	G(T,p,N) = -k_B T \ln \Delta_N(\beta, \gamma) .
\end{equation}

\section{One-dimensional nearest neighbour distribution in the isothermal-isobaric ensemble}

The results in sections~\ref{sec3.1} and~\ref{sec3.2} are general. Now we consider a one-dimensional fluid with nearest-neighbour interactions. In the isothermal-isobaric ensemble, whose probability distribution function is defined in~(\ref{II_probabilityfunction}), the evaluation of the nearest-neighbour probability distribution function can be done as follows (note that now the length $L$ plays the role of the volume $V$):
\begin{equation}\label{eq:iso1}
	p^{(1)}(r) \propto  \int^{\infty}_r \der L e^{-\beta p L} \int_{x_2}^L \der x_3 \int_{x_3}^L \der x_4 \cdots \int_{x_{N-1}}^L \der x_N  e^{-\beta\Phi_N(\textbf{r}^N)} .
\end{equation}

Now, it is possible to use~(\ref{NN_potential}) to rewrite (\ref{eq:iso1}) as
\begin{align}
	p^{(1)}(r) \propto &\enskip e^{-\beta\phi(r)} \int^{\infty}_r \der L e^{-\beta p L} \int_{x_2}^L \der x_3 e^{-\beta\phi(r_3)} \int_{x_3}^L \der x_4 e^{-\beta\phi(r_4)} \nonumber \\
	& \times \cdots \int_{x_{N-1}}^L \der x_N  e^{-\beta\phi(r_N)} e^{-\beta\phi(r_{N+1})},
\end{align}
where periodic boundary conditions have been applied and $r_{N+1}=L-r-r_3-r_4 \cdots -r_N$.

Using two changes of variable, the first one being $r_i=x_i-x_{i-1}$ and the second one being $L' = L-r$, the probability distribution function becomes
\begin{align}
p^{(1)}(r) \propto & \enskip e^{-\beta\phi(r)} \int^{\infty}_0 \der L' e^{-\beta p (L'+r)} \int_{0}^{L'} \der r_3 e^{-\beta\phi(r_3)} \int_{0}^{L'-r_3} \der r_4 e^{-\beta\phi(r_4)} \nonumber \\
& \times \cdots \int_{0}^{L'-r_3...-r_{N-1}} \der r_N  e^{-\beta\phi(r_N)} e^{-\beta\phi(r_{N+1})} .
\end{align}

Finally, after extracting a factor $e^{-\beta p r}$ from the first integral over $\der L'$, and since the remaining integrals do not depend on $r$, we obtain
\begin{equation}
	p^{(1)}(r) = K e^{-\beta\phi(r)} e^{-\beta p r},
\end{equation}
where $K$ is the normalization constant. The Laplace transform  of $p^{(1)}$ is
\begin{equation}
	\hat{P}^{(1)}(s)= K \hat{\Omega}(s+\beta p),
\end{equation}
where
\begin{equation}\label{eq:omegaLT}
	\hat{\Omega}(s)= \int_0^{\infty}\der r e^{-rs} e^{-\beta\phi(r)}
\end{equation}
is the Laplace transform of the pair Boltzmann factor $e^{-\beta\phi(r)}$.

Applying the normalization condition~(\ref{Pr:normalization2}) to determine $K$, we obtain
\begin{equation}
	K = \frac{1}{\hat{\Omega}(\beta p)}.
\end{equation}
Once the probability distribution function is known, we can apply~(\ref{G(s):calculation}) to compute the exact radial distribution function in the Laplace space:
\begin{align}\label{g(s)_formula}
\boxed{
	\hat{G}(s)= \frac{1}{n} \frac{\hat{\Omega}(s+\beta p)}{\hat{\Omega}(\beta p) - \hat{\Omega}(s+ \beta p)}.
}
\end{align}

\section{Equation of state}\label{section:equationofstate}

Through the determination of the radial distribution function and the probability distribution functions, it is possible to calculate certain thermodynamic quantities of the system. The most basic one is the equation of state, which relates the pressure, $p$, the density, $n$ and the temperature, $T$. To calculate it, we apply the condition that $g(r)$ must tend to $1$ for very large values of $r$:
\begin{equation}\label{FVT}
	\lim_{r \rightarrow \infty} g(r) = 1  \Rightarrow \lim_{s \rightarrow 0} s \hat{G}(s) = 1 ,
\end{equation}
 where the final value theorem has been used in the last step.

 Expanding $\hat{\Omega}(s+\beta p)$ in powers of $s$,
 \begin{align}
 	\hat{\Omega}(s+\beta p) &= \hat{\Omega}(\beta p) + \hat{\Omega}'(\beta p)  s + \frac{1}{2}\hat{\Omega}'' (\beta p)  s^2 + \cdots ,
 \end{align}
where $\hat{\Omega}'(s) = \partial \hat{\Omega}(s) / \partial s $ and $\hat{\Omega}''(s)=\partial^2\hat{\Omega} / \partial s^2$, we obtain
 	\begin{align}
 	\hat{G}(s) &= \frac{1}{n} \frac{\hat{\Omega}(\beta p) + \hat{\Omega}'(\beta p) s + \frac{1}{2}\hat{\Omega}'' (\beta p) s^2 + \cdots}{\hat{\Omega}(\beta p) - \hat{\Omega}(\beta p) - \hat{\Omega}'(\beta p) s - \frac{1}{2}\hat{\Omega}'' (\beta p) s^2 - \cdots} \nonumber \\
 	&= \frac{1}{n} \frac{\hat{\Omega}(\beta p) + \hat{\Omega}'(\beta p) s + \frac{1}{2}\hat{\Omega}'' (\beta p) s^2 + \cdots}{-s \left[ \hat{\Omega}'(\beta p) + \frac{1}{2}\hat{\Omega}'' (\beta p) s + \cdots \right]} .
 	\end{align}
Imposing condition~(\ref{FVT}),
 \begin{align}
 	\lim_{s \rightarrow 0} \frac{1}{n}& \frac{\hat{\Omega}(\beta p) + \hat{\Omega}'(\beta p) s + \frac{1}{2}\hat{\Omega}'' (\beta p) s^2 + \cdots}{- \hat{\Omega}'(\beta p) - \frac{1}{2}\hat{\Omega}'' (\beta p) s - \cdots }=1 .
 \end{align}

 Finally, the equation of state of the system can be evaluated as follows:
 \begin{equation}\label{EoS}
 \boxed{
 	n(p,T) = \frac{N}{V}=-\frac{\hat{\Omega}(\beta p)}{\hat{\Omega}'(\beta p)} .
 }
 \end{equation}

\section{Gibbs free energy}\label{section:gibbsfreeenergy}
 As discussed at the beginning of chapter~\ref{chap:isothermal_isobaric_ensemble}, the adequate thermodynamic potential in the isothermal-isobaric ensemble is the Gibbs free energy, $G(N,p,T)$.

 Using (\ref{gibbs:partials}) and~(\ref{EoS}), we obtain
 \begin{equation}
 	V = -N \frac{\hat{\Omega}'(\beta p)}{\hat{\Omega}(\beta p)} = -N \frac{\partial}{\partial (\beta p)} \ln \hat{\Omega}(\beta p) = -\frac{N}{\beta} \frac{\partial}{\partial p} \ln \hat{\Omega}(\beta p) = \left( \frac{\partial G}{\partial p} \right)_{N,T} ,
 \end{equation}
 which yields:
 \begin{equation}
 	G = -\frac{N}{\beta} \ln \hat{\Omega}(\beta p) + \Xi (N,T) ,
 \end{equation}
 where $\Xi(N,T)$ plays the role of an integration constant. In order to evaluate it, we impose the condition that, for very low pressures, the limit of the Gibbs free energy function must approach the one of the ideal gas, which is given by

\begin{equation}
G^{\text{ideal}} = \frac{N}{\beta} \ln (\beta p \Lambda),
\end{equation}
where
\begin{equation}
	\Lambda(\beta) \equiv \frac{h}{\sqrt{2 \pi m} / \beta}
\end{equation}
is the thermal de Broglie wavelength, $h$ being the Planck constant. Thus, we obtain
 \begin{equation}\label{eq:gibbs1}
 \lim_{p \rightarrow 0} \left[ -\frac{N}{\beta} \ln \hat{\Omega}(\beta p) \right] + \Xi (N,T) = \frac{N}{\beta} \ln (\beta p \Lambda).
 \end{equation}
 Taking into account the final value theorem and the general properties of limits and logarithms, the limit in the first term in~(\ref{eq:gibbs1}) becomes
 \begin{equation}
 	\lim_{p \rightarrow 0} \left[- \frac{N}{\beta} \ln \hat{\Omega}(\beta p) \right] = \frac{N}{\beta} \ln (\beta p),
 \end{equation}
 which yields
 \begin{equation}
 \Xi(N,T) = N k_B T \ln \Lambda .
 \end{equation}
Finally,
 \begin{equation}\label{eq:gibbs-omega}
 \boxed{
 	G(N,p,T)= N k_B T \ln \frac{\Lambda (\beta)}{\hat{\Omega}(\beta p; \beta)} .
 }
 \end{equation}

\section{Compressibility factor}\label{section:compressibilityfactor}

The compressibility factor is the ratio between the volume of a system to the volume of an ideal gas at the same conditions of temperature, $T$, pressure, $p$, and number of particles, $N$.

The equation of state of an ideal gas is given by
\begin{equation}
	pV^{\mathrm{ideal}}=k_B N T ,
\end{equation}
which yields a volume
\begin{equation}
	V^{\mathrm{ideal}}=\frac{N}{\beta p} .
\end{equation}
Therefore, the compressibility factor of a system with volume $V$ will be given by
 \begin{equation}\label{Z}
 	Z = \frac{V}{V^{\mathrm{ideal}}} = \frac{V}{N/\beta p} = \frac{\beta p}{n} .
 \end{equation}
 By definition, the compressibility factor of an ideal gas is always equal to $1$.

\section{Excess internal energy}\label{section:excessinternalenergy}

Using thermodynamic relations [see (\ref{Entropy_thermodynamics})], it is possible to write the energy of a system as
\begin{equation}\label{eq:thermodynamic_energy}
	E = G - pV + TS ,
\end{equation}
 where $G$ is defined in~(\ref{eq:gibbs-omega}) and the volume $V$ and the entropy $S$ can be calculated as follows:

 \begin{align}\label{eq:volume-omega}
 V &= \left( \frac{\partial G}{\partial p} \right)_{N,T} = \frac{\partial}{\partial p}\left[N k_B T \ln \frac{\Lambda (\beta)}{\hat{\Omega}(\beta p;\beta)} \right] \nonumber \\
 &= - N k_B T \Lambda(\beta) \frac{\partial}{\partial p} \ln \hat{\Omega}(\beta p;\beta) \nonumber \\
 & = - N k_B T \Lambda(\beta) \frac{\beta \hat{\Omega}'(\beta p;\beta)}{\hat{\Omega}(\beta p;\beta)} \nonumber \\
 &= -N \Lambda(\beta) \frac{\hat{\Omega}'(\beta p;\beta)}{\hat{\Omega}(\beta p;\beta)} ,
 \end{align}

 \begin{align} \label{eq:entropy-omega}
 	S &= \left( \frac{\partial G}{\partial T} \right)_{p,N} = \left( \frac{\partial G}{\partial \beta} \right)_{p,N} \left( \frac{\partial \beta}{\partial T} \right)_{p,N} = -\frac{1}{k_B T^2} \left( \frac{\partial G}{\partial \beta} \right)_{p,N} \nonumber \\
 	&=- \frac{N}{k_B T^2} \frac{\partial }{\partial \beta} \left[ \frac{1}{\beta} \ln \frac{\Lambda (\beta)}{\hat{\Omega}(\beta p;\beta)} \right] \nonumber \\
 	 &=-\frac{N}{k_B T^2} \frac{1}{\beta^2} \left[ \ln \frac{\Lambda(\beta)}{\hat{\Omega}(\beta p;\beta)} + \frac{1}{2} + \frac{\beta p \hat{\Omega}'(\beta p;\beta)-\beta \hat{\Upsilon}(\beta p;\beta)} {\hat{\Omega}(\beta p; \beta)} \right] \nonumber \\
 	 &= -N k_B \left[ \ln \frac{\Lambda(\beta)}{\hat{\Omega}(\beta p;\beta)} + \frac{1}{2} + \frac{\beta p \hat{\Omega}'(\beta p;\beta)-\beta \hat{\Upsilon}(\beta p;\beta)} {\hat{\Omega}(\beta p; \beta)} \right] ,
 \end{align}

 where
 \begin{equation}\label{eq:UpsilonDEF}
 	\hat{\Upsilon}(s) = - \frac{\partial \hat{\Omega}(s)}{\partial \beta} \equiv \int_0^{\infty} \der r e^{-rs}\phi(r) e^{-\beta\phi(r)} .
 \end{equation}

 Inserting (\ref{eq:gibbs-omega}),~(\ref{eq:volume-omega}) and~(\ref{eq:entropy-omega}) into~(\ref{eq:thermodynamic_energy}) we finally obtain
 \begin{equation}
 	E = N k_B T \left(\frac{1}{2} + \frac{\beta \hat{\Upsilon}(\beta p; \beta)}{\hat{\Omega}(\beta p; \beta)}\right) .
 \end{equation}
The excess internal energy of the system, that is, the total energy of the system, $E$ minus the kinetic part, can be written as

\begin{equation}\label{eq:internal_energy}
\boxed{
	\frac{\left\langle E \right\rangle^{\text{ex}}}{N} = \frac{\left\langle E \right\rangle}{N} - \frac{1}{2}k_BT = \frac{\hat{\Upsilon}(\beta p; \beta)}{\hat{\Omega}(\beta p; \beta)} .
}
\end{equation}

\chapter{Triangle well potential}\label{chap:tw}
 As a first application for one-dimensional systems, we will consider the triangle-well potential shown in Fig~\ref{trianglewell}, which is mathematically described by the function $\phi (r)$
 \begin{equation}\label{eq:TW}
\phi (r) =
 \begin{cases}
 \infty & \text{if $r\le \sigma$}, \\
 \displaystyle{-\frac{(r - \lambda) \epsilon}{\sigma - \lambda}} & \text{if $\sigma < r < \lambda$}, \\
 0 & \text{if $r \ge \lambda$},
 \end{cases}
 \end{equation}
 where $\sigma$ is the hard-core diameter of the particles, $\epsilon$ is the depth of the well and the condition that $\lambda \leq 2\sigma$  should be imposed in order for the interaction potential not to extend beyond the nearest neighbour.

The interaction potential $\phi (r)$ fulfils the conditions imposed in section~\ref{sec:nearest_neighbour_interaction} so it is possible to apply the exact results for one-dimensional systems obtained in chapter~\ref{chap:isothermal_isobaric_ensemble}.

Apart from its academic interest, the one-dimensional triangle-well potential is important because it exactly describes the effective colloid-colloid interaction in a colloid-polymer mixture in which the colloids are modelled by hard rods and the polymers are treated as ideal particles excluded from the colloids by a certain distance \cite{BE02}.

 The Laplace transform of the Boltzman factor [see (\ref{eq:omegaLT})] can be found to be

 \begin{equation}\label{eq:omegaTW}
 \hat{\Omega} (s) =- \frac{e^{-\lambda s }- e^{\beta \epsilon - s}}{\frac{\beta \epsilon}{\lambda - 1} + s} + \frac{e^{-\lambda s}}{s} ,
 \end{equation}
 where, for simplicity, we have taken $\sigma=1$ as the unit length. Equation~(\ref{eq:omegaTW}) can be conveniently rewritten the following way:
 \begin{equation}\label{eq:Omega}
\boxed{
 \hat{\Omega} (s)= \frac{X e^{-s}}{a-s} \left[\frac{a e^{-(\lambda-1)s}}{X s}-1 \right] ,
}
\end{equation}
where
 \begin{equation}\label{eq:Xa}
 	 X \equiv e^{\beta \epsilon}, \quad  	a \equiv -\frac{\beta \epsilon}{\lambda - 1}.
 \end{equation}
 The derivative of $\hat{\Omega}(s)$ with respect to $s$ is
 \begin{align}
 \hat{\Omega}'(s) = -\frac{e^{-\lambda s}}{s^2}+\frac{e^{-\lambda  s}-e^{\beta  \epsilon -s}}{\left(\frac{\beta  \epsilon }{\lambda -1}+s\right)^2}-\frac{e^{\beta  \epsilon -s}-\lambda  e^{-\lambda s}}{\frac{\beta  \epsilon }{\lambda -1}+s}-\frac{\lambda  e^{-\lambda s}}{s} .
 \end{align}
 If we rewrite this equation in terms of parameters $X$ and $a$, the result is
 \begin{equation}\label{eq:Omega'}
 \boxed{
 	\hat{\Omega}'(s)= -\frac{e^{-\lambda s}}{s^2}+\frac{e^{-\lambda  s}-Xe^{-s}}{\left(a-s\right)^2}+\frac{Xe^{-s}-\lambda  e^{-\lambda s}}{a-s}-\frac{\lambda  e^{-\lambda s}}{s} .
 }
 \end{equation}
 On the other hand,  the opposite of the derivative of $\hat{\Omega}(s)$ with respect to $\beta$ [see~(\ref{eq:UpsilonDEF})] is
\begin{equation}
 	  \hat{\Upsilon} (s) = -\frac{\epsilon  e^{\beta \epsilon -s}}{\frac{\beta \epsilon }{\lambda -1}+s}-\frac{\epsilon  \left(e^{- \lambda s}-e^{\beta \epsilon -s}\right)}{(\lambda -1) \left(\frac{\beta \epsilon }{\lambda -1}+s\right)^2} ,
\end{equation}
that can be also written as a function of the parameters $X$ and $a$:
 \begin{equation}\label{eq:Upsilon}
 \boxed{
	\hat{\Upsilon}(s) = \frac{\epsilon}{a-s} \left[Xe^{-s} - \frac{e^{-\lambda s}-Xe^{-s}}{(\lambda-1)(a-s)}\right].
}
 \end{equation}

\section{Thermodynamic quantities}
Once the functions $\hat{\Omega}(s)$, $\hat{\Omega}'(s)$ and  $\hat{\Upsilon}(s)$ are known, it is possible to calculate the relevant thermodynamic properties of the liquid as shown in sections \ref{section:equationofstate}--\ref{section:excessinternalenergy}. Here we focus on the compressibility factor and the excess internal energy.

\subsection*{Compressibility factor}

According to~(\ref{EoS}), and taking into account~(\ref{eq:Omega}) and~(\ref{eq:Omega'}), the number density of the system is given by
\begin{equation}\label{eq:density-tw}
	\frac{1}{n} = 1+\frac{1}{\beta p-a}+\frac{a e^{\beta p} \left[(\lambda -1)\beta p+1 \right]}{\beta p \left(a e^{\beta p}-\beta p\hspace{1pt} X \hspace{1pt} e^{\beta p\lambda}\right)} .
\end{equation}
In order to evaluate the compressibility factor of a system [see~(\ref{Z})], it is necessary to compute $\beta p(n)$. As this cannot be done analytically from~(\ref{eq:density-tw}), numerical methods have been used to calculate the value of $\beta p$ for every pair $n$ and $\beta$.

\begin{figure}[htpb]
	\centering
	\includegraphics[scale=0.9]{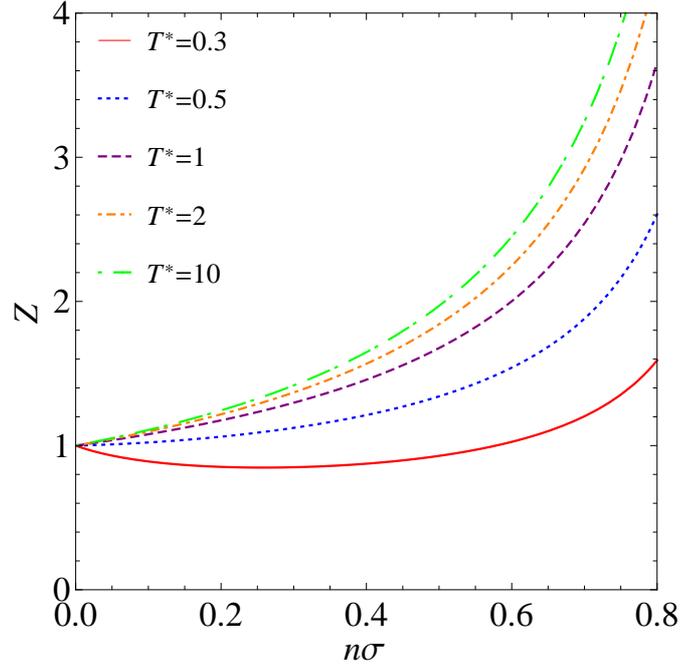}
	\caption{Density dependence of the compressibility factor $Z=\beta p / n$ of the one-dimensional triangle-well potential for several reduced temperatures $T^* \equiv k_BT/\epsilon$ at $\lambda=1.4\sigma$.}
	\label{fig:well_z}
\end{figure}

Figure~\ref{fig:well_z} shows the density dependence of the compressibility factor for several temperatures in the representative case $\lambda=1.4\sigma$. As expected, the compressibility factor, $Z$, tends to $1$ as density approaches zero (regardless of temperature), since the properties of the system for very low densities must coincide with those of the ideal gas.

On the other had, it is worth noticing that, if the temperature is low enough (as it is when $T^*=0.3$) the compressibility factor decreases when we increase the density until it reaches a minimum. This initial decrease is due to the influence of the attractive part of the potential. However, for relatively high temperatures, the compressibility factor always increases as density gets higher.

\subsection*{Excess internal energy}

The excess internal energy [see~(\ref{eq:internal_energy})] can be evaluated by substituting the expressions for $\hat{\Omega}(s)$ and $\hat{\Upsilon}(s)$ from (\ref{eq:Omega}) and (\ref{eq:Upsilon}), respectively, giving the following expression
\begin{equation}\label{eq:energy-tw}
\frac{\left\langle E \right\rangle^{\text{ex}}}{N}= -\beta p \frac{X e^{\beta p \lambda } [-a\lambda+a+\beta p (\lambda -1)-1]+e^{\beta p}}{(\lambda -1) (a-\beta p) \left(a e^{\beta p}-\beta p X e^{\beta p \lambda }\right)} .
\end{equation}

\begin{figure}[htpb]
	\centering
	\includegraphics[scale=0.9]{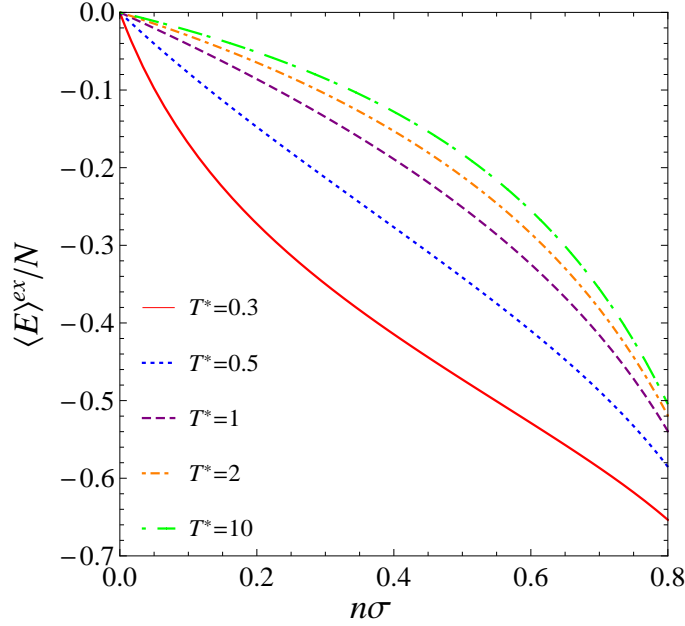}
	\caption{Density dependence of the excess internal energy $\left\langle E \right\rangle^{ex}/N$ of the one-dimensional triangle-well potential for several reduced temperatures $T^* \equiv k_BT/\epsilon$ at $\lambda=1.4\sigma$.}
	\label{fig:well_energy}
\end{figure}

From Fig.~\ref{fig:well_energy} it is possible to see how the internal energy per particle decreases as density and temperature decrease. Because we have a triangle-well, particles in the system will attract each other whenever they get close enough, lowering their potential energy. This means that, for high densities, the internal energy of the system will be lower than for smaller densities.

\section{Radial distribution function}\label{sec:rdf}

In order to calculate the radial distribution function, $g(r)$, it is necessary to calculate its Laplace transform $\hat{G}(s)$ first. The function $\hat{G}(s)$ can be evaluated analytically through (\ref{g(s)_formula}) but, in order to perform the inverse Laplace transform, it is more convenient to rewrite it in a series form:
\begin{equation}
	\hat{G}(s)= \frac{1}{n} \frac{\hat{\Omega}(s+\beta p)}{\hat{\Omega}(\beta p) - \hat{\Omega}(s+ \beta p)} =
	\frac{1}{n} \sum_{l=1}^{\infty}      \left[ \frac{\hat{\Omega}(s+\beta p)}{\hat{\Omega}(\beta p)} \right]^l .
\end{equation}

The value of $\hat{\Omega}(\beta p)$ is constant with respect to $s$ so, in order to compute the inverse Laplace transform, it is only necessary to focus on the numerator, which is obtained from (\ref{eq:Omega}) as
\begin{align}
	\left[\hat{\Omega}(s + \beta p)\right]^l &=
	\frac{e^{-\beta p l} e^{-l s} X^l }{(a-\beta p-s)^l} \sum _{h=0}^l
{l \choose h}	
	(-1)^{l-h} \left(\frac{a}{X}\right)^{h} e^{-\beta p (\lambda -1) h} \frac{e^{-h(\lambda -1) s}}{(\beta p+s)^{h}}\nonumber \\
	&\equiv \sum_{h=0}^{l} C_{lh} \hat{F}_{lh}(s) ,
\end{align}
where we have called
\begin{equation}\label{C1}
	C_{lh} \equiv \frac{l! (-1)^{l-h} \left(\frac{a}{X}\right)^h e^{-\beta p h (\lambda -1)} e^{-\beta p l} X^l}{h! (l-h)!}
\end{equation}
and
\begin{equation}\label{Fs}
	\hat{F}_{lh}(s) = \frac{e^{-l s} e^{-h (\lambda -1) s}}{(\beta p+s)^h (a-\beta p-s)^l} .
\end{equation}

It is not difficult to observe that the quantity $C_{lh}$ is a constant with respect to $s$ so we only need to compute the inverse Laplace transform of $\hat{F}_{lh}(s)$. Therefore, the inverse Laplace transform of $\hat{G}(s)$ can be calculated as follows:

\begin{align}
\begin{split}
g(r) &= \mathcal{L}^{-1}[\hat{G}(s)] = \frac{1}{n}\sum_{l=1}^{\infty} \frac{\mathcal{L}^{-1} \left[(\hat{\Omega}(s+\beta p))^l\right]}{\left[\hat{\Omega}(\beta p)\right]^l}= \frac{1}{n} \sum_{l=1}^{\infty} \frac{\sum_{h=0}^l C_{lh} \mathcal{L}^{-1}\left[F_{lh}(s)\right]}{\left[\hat{\Omega}(\beta p)\right]^l},
\end{split}
\end{align}
where $\mathcal{L}^{-1}[\cdots]$ denotes the inverse Laplace transform. The only thing left to do is to compute $\mathcal{L}^{-1}[\hat{F}_{lh}(s)]=f_{lh}(r)$. To do that, we note that, after a lengthy calculation \cite{NotesForAna}, (\ref{Fs}) can be rewritten as
\begin{equation}
	\hat{F}_{lh}(s) = \sum _{n_1=1}^h \frac{\binom{h+l-n_1-1}{l-1}}{a^{h+l-n_1}} \frac{e^{-l s} e^{-h (\lambda -1) s}}{(\beta p+s)^{n_1}}+ \sum _{n=1}^l
	\frac{\binom{h+l-n-1}{h-1}}{a^{h+l-n}}
	\frac{ e^{-l s} e^{-h (\lambda -1) s}}{(a-\beta p-s)^n } .
\end{equation}
Now, taking into account the mathematical property \cite{Abramowitz}
\begin{equation}
	\mathcal{L}^{-1} \left[\frac{e^{-\alpha s}}{(s+\tau)^l}\right] = \frac{(r-\alpha)^{l-1}}{(l-1)!}e^{-\tau(r-\alpha)}\Theta (r-\alpha),
\end{equation}
where $\Theta(\cdots)$ denotes the Heaviside step function, one gets
\begin{align}\label{f(r)}
	f_{lh}(r) =&\sum _{n_1=1}^h \frac{\binom{h+l-n_1-1}{l-1}}{a^{h+l-n_1}} \frac{ \left[r-l-h (\lambda -1) \right]^{n_1-1} e^{-\beta p[r-h (\lambda -1)-l]}}{(n_1-1)!}\Theta (r-l-h (\lambda -1)) \nonumber \\
	&+\sum _{n=1}^l
	\frac{\binom{h+l-n-1}{h-1}}{a^{h+l-n}}
    \frac{\left[r-l-h (\lambda -1)\right]^{n-1} e^{(a-\beta p) [-h (\lambda -1)-l+r]}}{(n-1)!}(-1)^{-n} \nonumber \\
    & \times \Theta (r-l-h (\lambda -1)) .
\end{align}

Finally, the radial distribution function of the system, $g(r)$, can be calculated as follows:
\begin{equation}\label{eq:rdf}
\boxed{
	 g(r)=\frac{1}{n} \sum_{l=1}^{\infty}  \frac{\sum_{h=0}^l C_{lh} f_{lh}(r)}{\left[\hat{\Omega}(\beta p)\right]^l} ,
	}
\end{equation}
 where $C_{lh}$ and $f_{lh}(r)$ are given by (\ref{C1}) and (\ref{f(r)}), respectively.
\begin{figure}[htpb]
	\centering
	\includegraphics[scale=0.85]{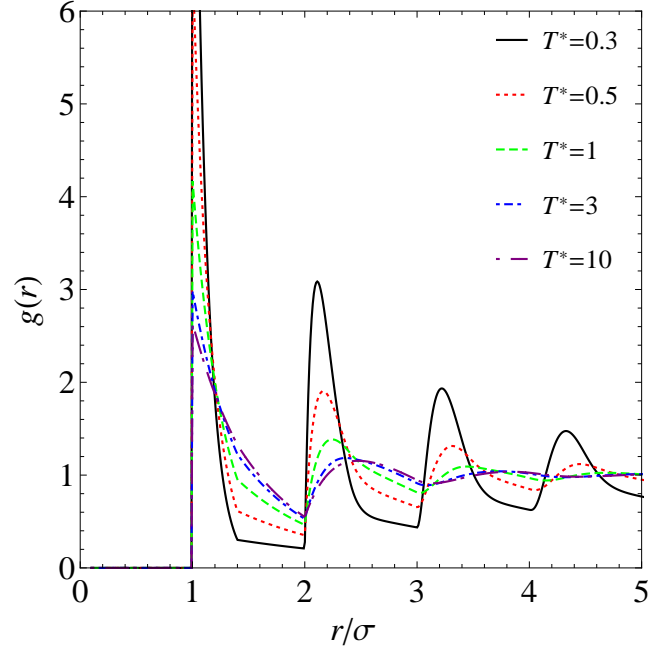}
	\caption{Plot of the radial distribution function of the one-dimensional triangle-well fluid for several temperatures at $n^*=0.6$ and $\lambda = 1.4\sigma$.}
	\label{fig:gr_T}
\end{figure}
It is important to note that, although~(\ref{eq:rdf}) contains an infinite number of terms, thanks to the structure of the Heaviside step function, only the terms up to $l=l_{\text{max}}$ are needed if we are only interested in $g(r)$ until $r=l_{\text{max}}+1$.

Figure~\ref{fig:gr_T} shows that, for lower temperatures, the peaks of the radial distribution function (which mark the most probable position of the particles around a given reference one) are higher and tend to be closer to $r/\sigma=2,3,\ldots$ than for higher temperatures, in which the radial distribution function quickly tends to one. This means that, for lower temperatures, there exists a short-range order in the positions of the particles that disappears when the distance to the origin particle is high enough. On the contrary, for high temperatures the radial distribution function quickly goes to one, meaning that this short-range order tends to disappear if the temperature is high enough

Also, as expected, the radial distribution function is zero for $r < \sigma$ since two particles cannot approach their centres closer than a distance $r=\sigma$.

\begin{figure}[htpb]
	\centering
	\includegraphics[scale=0.85]{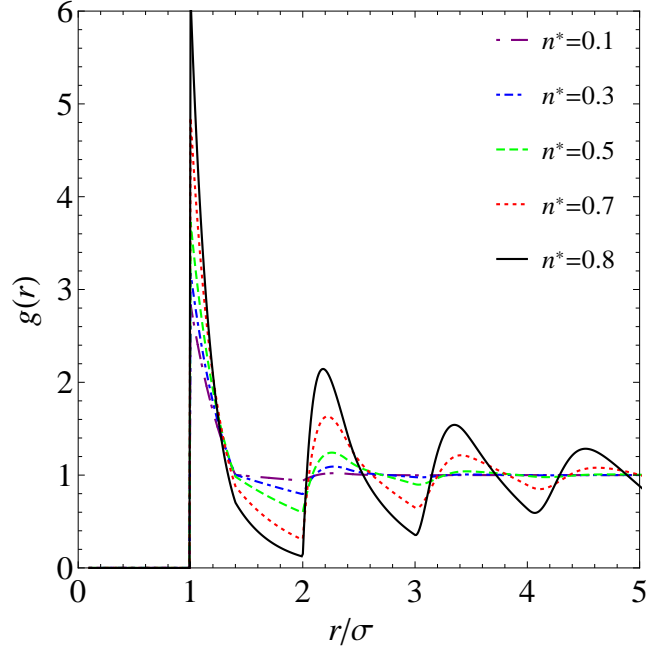}
	\caption{Plot of the radial distribution function of the one-dimensional triangle-well fluid for several densities at $T^*=1$ and $\lambda=1.4\sigma$.}
	\label{fig:gr_n}
\end{figure}

In Fig.~\ref{fig:gr_n}, we can observe that when density is low, the average distance between the particles is high so the system behaves almost like an ideal gas (except for the exclusion due to the condition that two particles cannot be closer than a distance $r=\sigma$) so the radial distribution function quickly tends to one.

For higher densities, particles tend to pack in a fixed structure at distances marked by the maxima in the radial distribution function.

\section{Structure factor}

The relation between the static structure factor $\tilde{S}(k)$ and the total correlation function $\tilde{h}(k)$ is described in (\ref{eq:Sk}) and thus, in order to calculate the structure factor, it is first necessary to obtain $\tilde{h}(k)=\mathcal{F}\left[h(r)\right]$, where $\mathcal{F}\left[\cdots\right]$ denotes de Fourier transform.

The Laplace transform of the total correlation function is
\begin{equation}
	\hat{H}(s) = \int_o^{\infty} \der r e^{-rs} h(r) =\int_o^{\infty} \der r e^{-rs} \left[g(r)-1\right] = \hat{G}(s)-\frac{1}{s}.
\end{equation}
Thanks to the properties of the Fourier and Laplace transform, $\tilde{h}(k)$ can be finally obtained as
\begin{equation}
	\tilde{h}(k) = \left[\hat{H}(s) + \hat{H}(-s)\right]_{s=\imath k}=\left[\hat{G}(s) + \hat{G}(-s)\right]_{s=\imath k} = 2 \text{Re}\left[\hat{G}(\imath k)\right] ,
\end{equation}
where $\text{Re}\left[z\right]$ denotes the real part of the complex number $z$ and the analytical expression for $\hat{G}(s)$ can be obtained from~(\ref{g(s)_formula}).

\begin{figure}[htpb]
	\centering
	\includegraphics[scale=0.85]{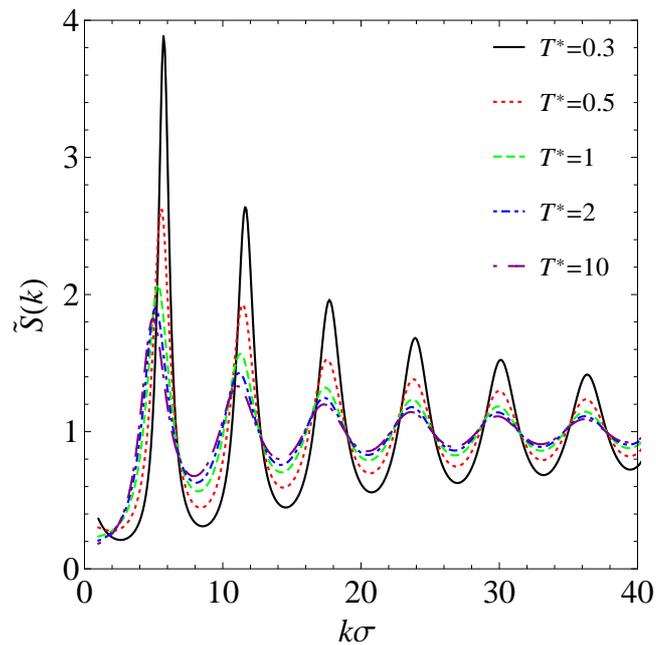}
	\caption{Plot of the structure factor of the one-dimensional triangle-well potential for several temperatures at $n^*=0.6$ and $\lambda = 1.4 \sigma$.}
	\label{fig:sk_T}
\end{figure}

\begin{figure}[htpb]
	\centering
	\includegraphics[scale=0.85]{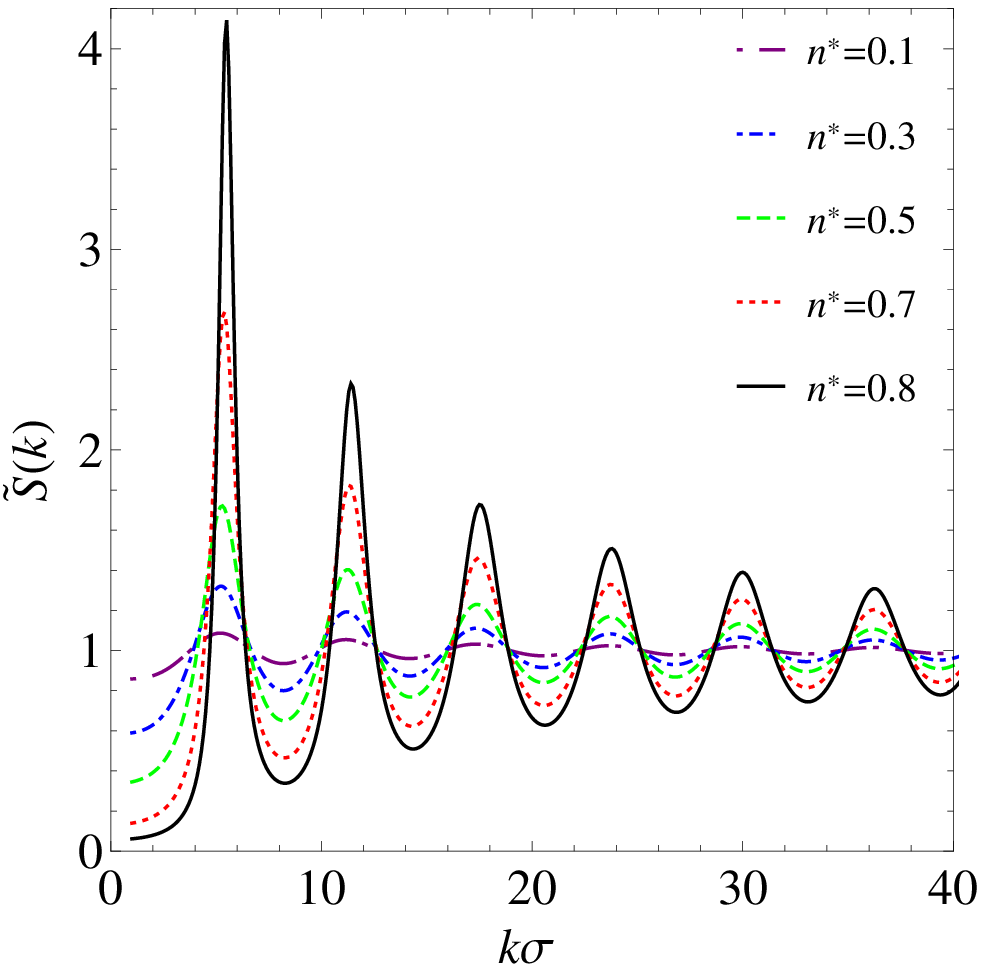}
	\caption{Plot of the structure factor of the one-dimensional triangle-well potential for several densities at $T^*=1$ and $\lambda = 1.4 \sigma$.}
	\label{fig:sk_n}
\end{figure}

Figure~\ref{fig:sk_T} shows the structure factor for several temperatures, where it can be seen that for lower temperatures the peaks in $S(k)$ are higher due to the fact that the scattering intensity is directly related to the structure of the liquid. For higher temperatures (that is, as the system gets more disordered) the structure factor tends to be closer to one, which is the value of the structure factor of an ideal gas.

In Fig.~\ref{fig:sk_n} we can clearly see that, for systems approaching the ideal gas behaviour (systems with very low densities), the structure factor is practically equal to one, meaning that there will be no privileged directions in the scattered radiation.

\section{Direct correlation function}\label{sec:dcf}

\begin{figure}[htpb]
	\centering
	\includegraphics[scale=0.85]{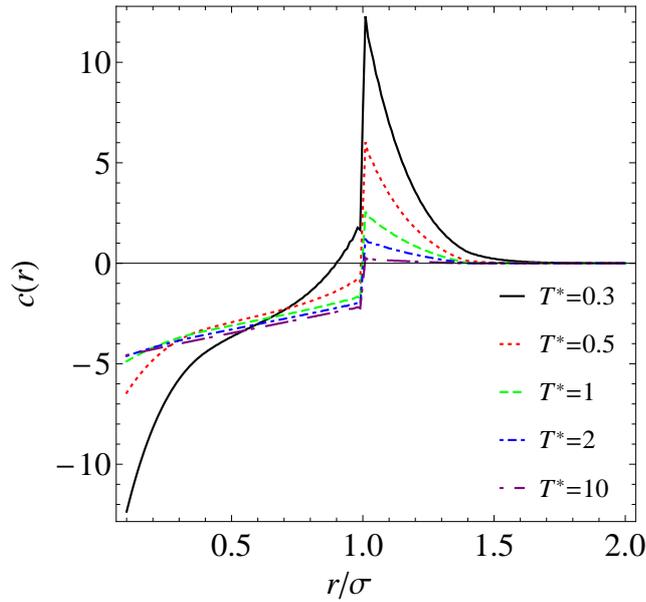}
	\caption{Plot of the direct correlation function of the one-dimensional triangle-well fluid for several temperatures at $n^*=0.6$ and $\lambda = 1.4 \sigma$.}
	\label{fig:cr_T}
\end{figure}

Thanks to (\ref{eq:relationsCF}) it is also possible to calculate the Fourier transform of the direct correlation function, $\tilde{c}(k)$,  through $\tilde{h}(k)$. Then, a numerical inverse Fourier transform yields $c(r)$. Figure~\ref{fig:cr_T} displays the direct correlation function for several temperatures.

We can observe that $c(r)$ presents a jump at $r=\sigma$, right when the potential is discontinuous and that it is generally small in the region $r > \lambda$, that is, beyond the range of interaction potential. This shows that, as expected, the direct correlation function is shorter-ranged than the total one because we are erasing the indirect part of the correlation.

Also, the correlation between the particles that are inside the range of interaction  of the potential $(r<\lambda)$ is bigger when temperatures are lower and decreases when temperatures are higher. In general, for all temperatures the correlation function reaches a maximum at $r=\sigma$, when the depth of the triangle-well is maximum.

\begin{figure}[htpb]
	\centering
	\includegraphics[scale=0.85]{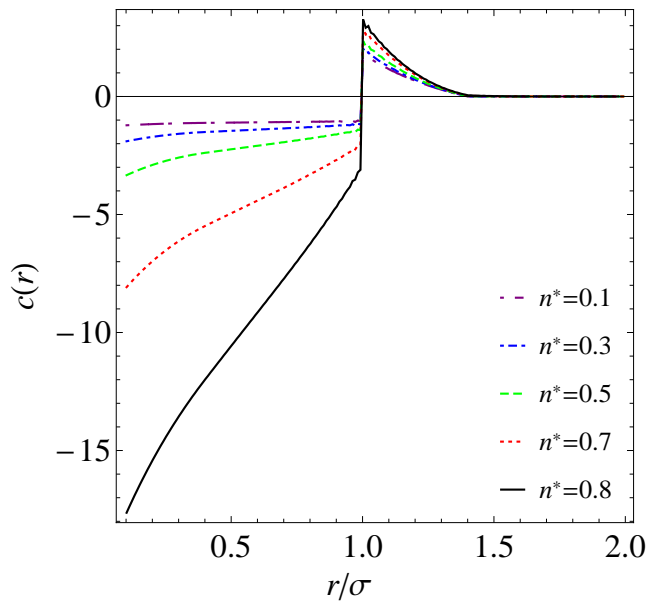}
	\caption{Plot of the direct correlation function of the one-dimensional triangle-well fluid for several densities at $T^*=1$ and $\lambda = 1.4 \sigma$.}
	\label{fig:cr_n}
\end{figure}

Figure~\ref{fig:cr_n} shows the direct correlation function for several densities at $T^*=1$. For all densities, the direct correlation function presents a jump at $r=\sigma$. This jump is bigger for higher densities since, in these cases, $c(r)$ is more negative in the region $r<\sigma$ but reaches higher values in the region $\sigma<r<\lambda$. On the other hand, regardless of the density, $c(r)$ goes quickly to zero for $r>\lambda$.

\section{Test of approximate closures for the direct correlation function}

We recall that the direct correlation function was defined by the Ornstein-Zernike equation (\ref{OZ-equation}). But this equation is not a closed one so, unless the exact solution of the problem is known, its is necessary to find an approximate closure of the form $c(r)=c_{\text{approx}}[h(r)]$ in order for the Ornstein-Zernike equation to become an approximate closed integral equation that can be solved:

\begin{equation}
	h(r) = c_{\mathrm{approx}}[h(r)] + n \int \der \textbf{r}' c_{\mathrm{approx}}[h(r')] h(|\textbf{r}-\textbf{r}'|).
\end{equation}

The two prototype closures are the Percus--Yevick approximation (PY) \cite{PY} and the hypernetted-chain (HNC) \cite{HNC}. The latter closure is
\begin{equation}
	c_{\mathrm{HNC}}(r) = g(r) -1-\ln[g(r)]-\beta\phi(r).
\end{equation}
The Percus--Yevick approximation consists in taking the following equation for $c(r)$
\begin{equation}
	c_{\mathrm{PY}}(r) = g(r) \left[1-e^{\beta\phi(r)}\right].
\end{equation}
It is worth noticing that, due to the fact that the potential is infinite for $r<\sigma$, both approximate equations do not define $c(r)$ in that region. Also, the interaction potential is zero in the region $r>\lambda$ and, therefore, the approximate quantity $c_{\text{PY}}(r)$ will also equal zero in that region.

Since we have already obtained the exact solution for the radial distribution function, $g(r)$, and the direct correlation function, $c(r)$, in sections~\ref{sec:rdf} and~\ref{sec:dcf}, respectively, it is now possible to check which one of the two approximate closures deviates less from the exact solution. In order to do that, we will compare them for different values of the parameter $\lambda$ as well as for different densities and temperatures.

\subsection{Behaviour for high temperatures}
\begin{figure}[htbp]
	\centering
	\subfigure[$n^*=0.2$.]{\includegraphics[height=7.3cm]{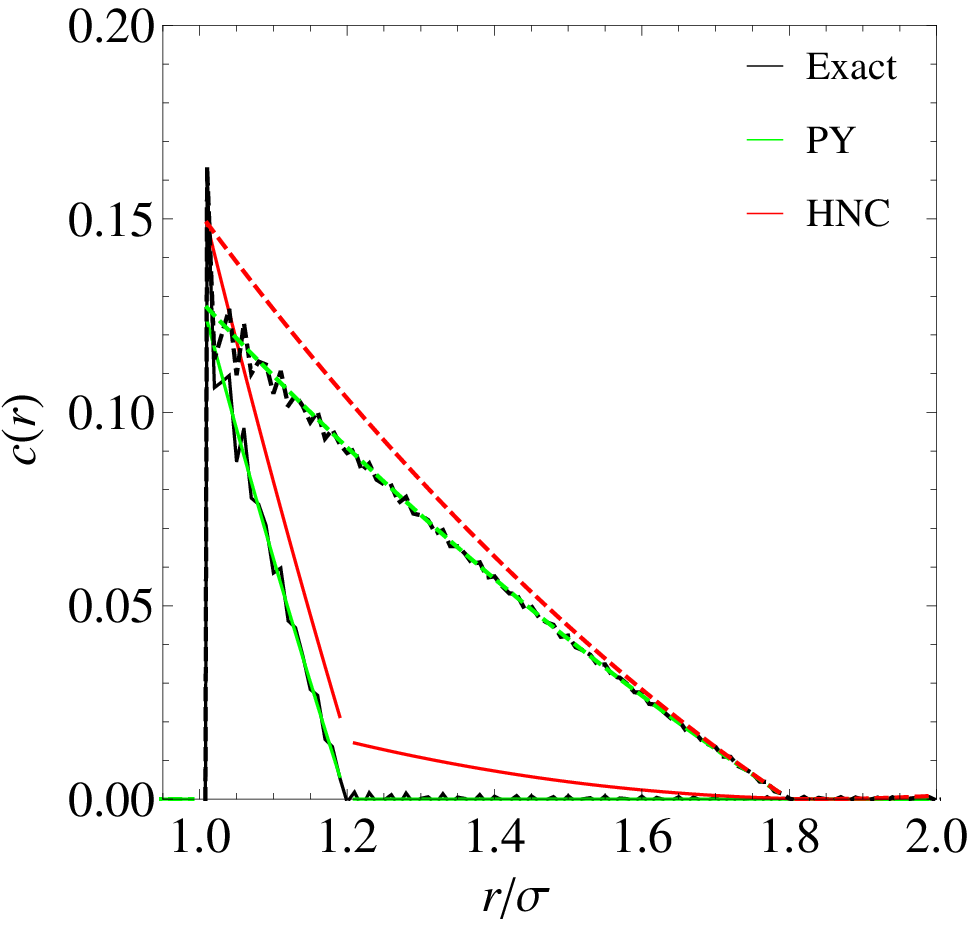}}
	\subfigure[$n^*=0.7$.]{\includegraphics[height=7.3cm]{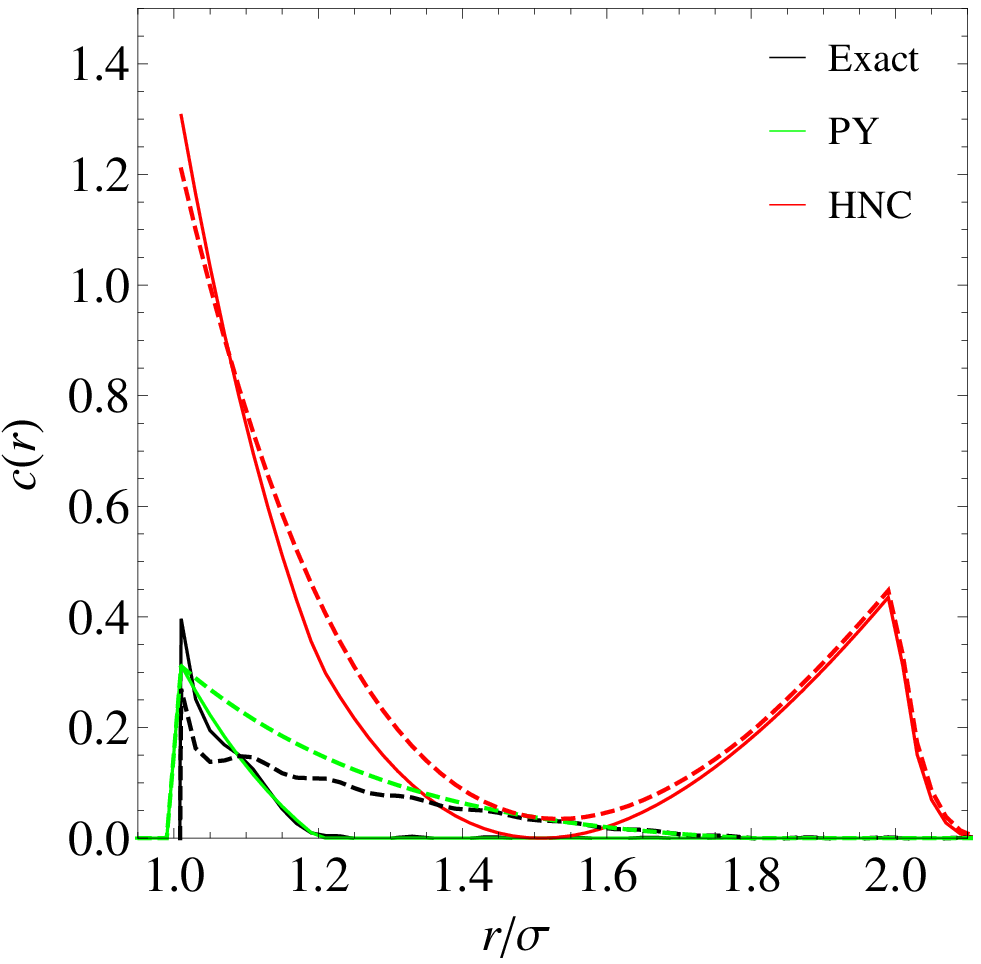}}
	\caption{Plot of the direct correlation function $c(r)$ and the PY and HNC approximate values for $c(r)$ when $T^*=10$ and (a) $n^*=0.2$, (b) $n^*=0.7$. The dotted lines represent the direct correlation function when $\lambda=1.8\sigma$ and the lines represent the direct correlation function when $\lambda=1.2\sigma$.} \label{fig:crapprox_tt}
\end{figure}

Figure~\ref{fig:crapprox_tt} shows how close the approximations HNC and PY are to the analytical solution for $c(r)$. For low densities, both approximations work fairly well but the PY approximation always stays closer to the analytical solution than the HNC one. When we take larger values for $\lambda$, the PY approximation is still similar to the analytical solution but the HNC approximation is quite different to the supposed result.

\subsection{Behaviour for low temperatures}

\begin{figure}[htbp]
	\centering
	\subfigure[$n^*=0.2$]{\includegraphics[height=7.55cm]{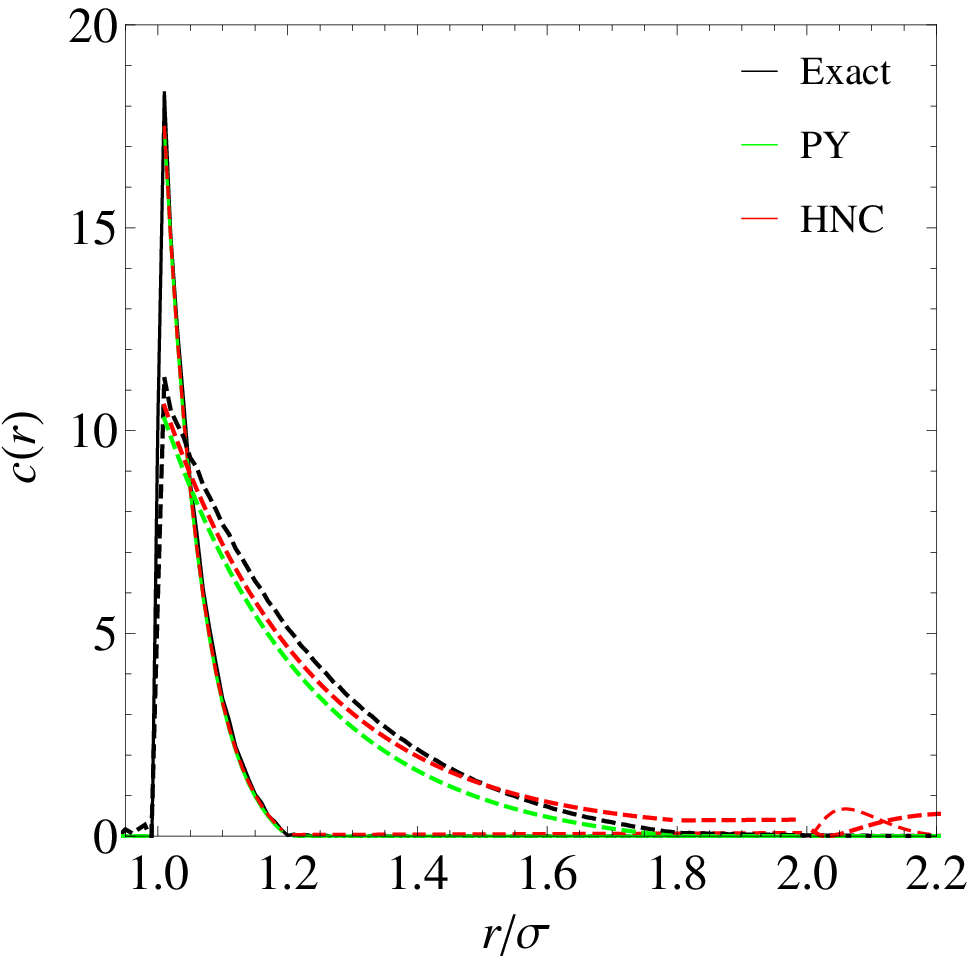}}
	\subfigure[$n^*=0.7$]{\includegraphics[height=7.3cm]{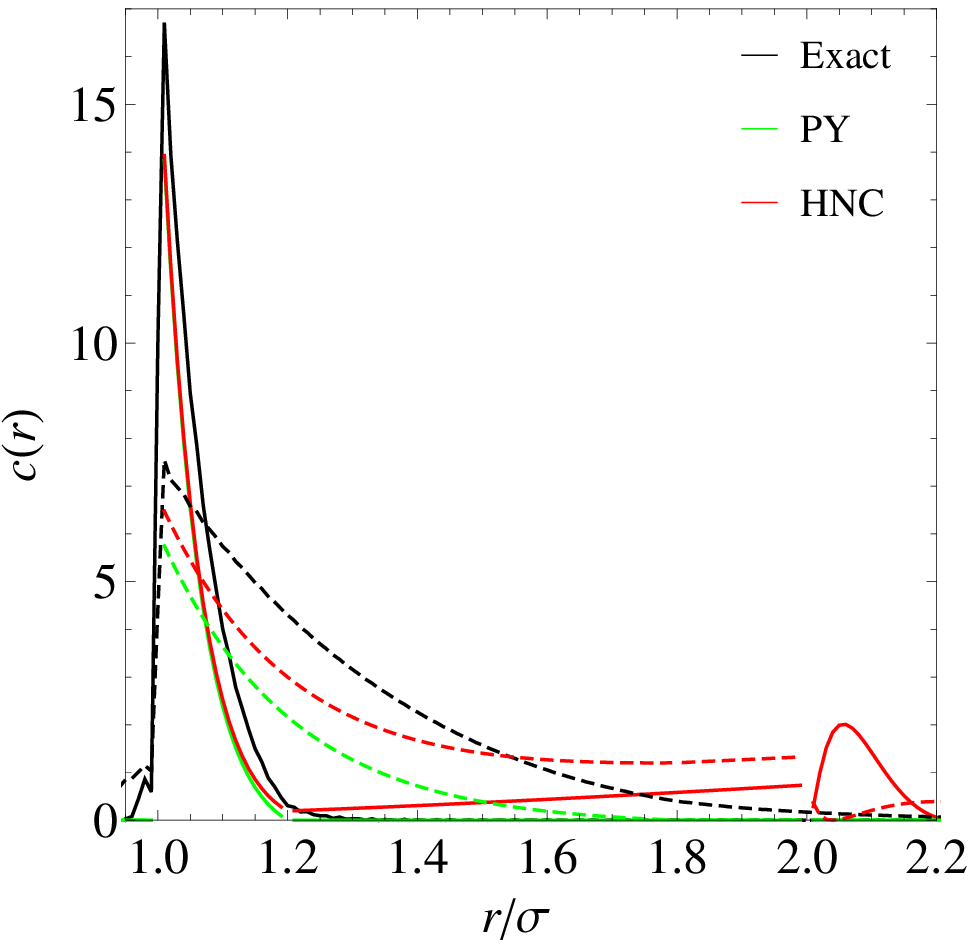}}
	\caption{Plot of the direct correlation function $c(r)$ and the PY and HNC approximate values for $c(r)$ when $T^*=0.3$ and (a) $n^*=0.2$, (b) $n^*=0.7$. The dotted lines represent the direct correlation function when $\lambda=1.8\sigma$ and the lines represent the direct correlation function when $\lambda=1.2\sigma$.} \label{fig:crapprox_t}
\end{figure}

Figure~\ref{fig:crapprox_t} shows how approximate $c_{\mathrm{PY}}(r)$ and $c_{\mathrm{HNC}}(r)$ are to the analytical solution when the temperature of the system is low. For low densities, both expression approximate well to the analytical solution but the PY approximation stays closer to it in the range $r>2\sigma$. When the density is higher, the HNC approximation behaves better for $r \simeq \sigma$ but behaves worse when $r\simeq 2\sigma$ because it starts oscillating instead of going to zero.

In general, it is possible to say that the PY approximation stays closer to the exact solution for all the different possibilities tried out. For low densities, both approximations are quite close to the real solution but, when we increase density, the HNC approximation can be quite different from the exact result, while the PY stays closer to it.

\section{Asymptotic behaviour of the radial distribution function}

Once the function $\hat{G}(s)$ is known, it is possible to calculate $g(r)$ by means of the inverse Laplace transform \cite{Laplace}:
\begin{equation}\label{eq:InverseLaplace}
	g(r)=\frac{1}{2 \pi \imath}\lim_{T \rightarrow \infty}\int_{\gamma - \imath T}^{\gamma + \imath T} \der s e^{st}\hat{G}(s)  ,
\end{equation}
where the integration is done along the vertical line $\text{Re}(s) = \gamma$ in the complex plane such that all singularities of $\hat{G}(s)$ must lie to the left. That is, $\gamma$ must be greater than the real part of all singularities of $\hat{G}(s)$. In this situation, the integral (\ref{eq:InverseLaplace}) can be calculated as

\begin{equation}\label{eq:InverseRes}
	\frac{1}{2 \pi \imath}\lim_{T \rightarrow \infty}\int_{\gamma - \imath T}^{\gamma + \imath T}  \der s e^{st}\hat{G}(s) = \sum_j \text{Res}\left[e^{s_jr}\hat{G}(s_j)\right]=\sum_j e^{s_jr} \text{Res}\left[\hat{G}(s_j)\right] ,
\end{equation}
where $s_j$, $j=1,2, \ldots$ are the poles of the function $\hat{G}(s)$ and $\text{Res}\left[\cdots\right]$ is the residue. Finally, we arrive at the important result:
\begin{equation}\label{eq:gr_res}
	\boxed{
     g(r)=\sum_j e^{s_jr} \text{Res}\left[\hat{G}(s_j)\right] ,
}
\end{equation}
from which it is worth noticing that $\text{Re}[s_j]<0$ for all poles of $\hat{G}(s)$; if it were not so, the integral (\ref{eq:InverseRes}) will diverge for very large values of $r$.
\subsection*{Poles of $\hat{G}(s)$}

Also, taking into account (\ref{g(s)_formula}), the poles of $\hat{G}(s)$, denoted by $s_j$, will satisfy the following equation:
\begin{equation}\label{eq:res0}
\hat{\Omega}(\beta p) = \hat{\Omega}(s_j+\beta p),
\end{equation}
from which it can be easily checked that there is always a pole at $s_j=0$. Condition (\ref{eq:res0}) can be also written as a set of two equations if we separate the real and imaginary parts:
\begin{gather}\label{eq:complexpole}
\begin{cases}
\text{Re}\left[\hat{\Omega}(s_j+\beta p)\right] = \hat{\Omega}(\beta p), \\
\text{Im}\left[\hat{\Omega}(s_j+\beta p)\right] = 0, \\
\end{cases}
\end{gather}
where $\text{Re}\left[z\right]$ and $\text{Im}\left[z\right]$ represent the real and imaginary part of the complex number $z$, respectively.

\subsection*{Residues of $\hat{G}(s)$}

In order to find the analytic expression for the residues of $\hat{G}(s_j)$, let us write the function $\Omega(s+\beta p)$ as a power series expansion,
\begin{equation}\label{eq:res1}
\hat{\Omega}(s+\beta p) = \hat{\Omega}(s_j+\beta p) + \hat{\Omega}'(s_j + \beta p)(s-s_j) + O\left[(s-s_j)^2\right] .
\end{equation}
Now, the residue of a simple pole of a complex function is evaluated by taking the following limit:
\begin{align}\label{eq:res2}
\begin{split}
\text{Res}\left[\hat{G}(s_j)\right] &= \lim_{s \rightarrow s_j} (s-s_j) \hat{G}(s) = \frac{1}{n}\lim_{s \rightarrow s_j} (s-s_j) \frac{\hat{\Omega}(s+\beta p)}{\hat{\Omega}(\beta p)-\hat{\Omega}(s+\beta p)},
\end{split}
\end{align}
which, taking into account (\ref{eq:res0}) and (\ref{eq:res1}), yields
\begin{equation}
\text{Res}\left[\hat{G}(s_j)\right] =- \frac{1}{n}\frac{\hat{\Omega}(\beta p)}{\hat{\Omega}'(s_j+\beta p)} ,
\end{equation}
that can be expressed in a more convenient way by means of the equation of state of the system (\ref{EoS})
\begin{equation}\label{residues}
\boxed{
	\text{Res}\left[\hat{G}(s_j)\right] =\frac{\hat{\Omega}'(\beta p)}{\hat{\Omega}'(s_j+\beta p)} .
}
\end{equation}

Let us now call $s_0$ to the pole of $\hat{G}(s)$ located at the origin, that is, $s_0=0$. In this case, as it can be seen in (\ref{residues}), the residue will be
\begin{equation}\label{eq:res_0}
\text{Res}\left[\hat{G}(s_0)\right] = 1.
\end{equation}

\subsection*{Asymptotic behaviour}

If we are interested in the behaviour of $g(r)$ when $r \rightarrow \infty$, then, taking into account (\ref{eq:gr_res}) and (\ref{eq:res_0}), the following relation holds:
\begin{equation}
\lim_{r \rightarrow \infty} g(r) = \lim_{r \rightarrow \infty} \sum_j e^{s_jr} \text{Res}\left[\hat{G}(s_j)\right] = 1+e^{s_kr} \text{Res}\left[\hat{G}(s_k)\right] ,
\end{equation}
where $s_k$ is the pole of $\hat{G}(s)$ with the real part closest to the origin. In order to calculate this pole, we need to distinguish whether this pole is real or a pair of complex conjugates since the  asymptotic behaviour of the radial distribution function will depend on this.

\begin{itemize}
	\item The closest pole to the origin is a pair of complex conjugates, $s_1=-\kappa + i\omega$ and $s_2=-\kappa - i \omega$. In that case,
	\begin{equation}\label{eq:poles_cc}
		\lim_{r \rightarrow \infty} g(r) = 1+e^{s_1r} \text{Res}\left[\hat{G}(s_1)\right]+e^{s_2r} \text{Res}\left[\hat{G}(s_2)\right] ,
	\end{equation}
	where $\text{Res}\left[\hat{G}(s_1)\right] = |A| e^{\imath \delta}$ and $\text{Res}\left[\hat{G}(s_2)\right] = |A| e^{-i \delta}$. Then, it is possible to rewrite (\ref{eq:poles_cc}) as
	\begin{align}\label{eq:oscillatory}
		\lim_{r \rightarrow \infty} g(r) &= 1 + |A| e^{-\kappa} \left[ e^{\imath(\omega r + \delta)}+e^{-\imath(\omega + \delta)}\right] \nonumber \\
		&= 1+2|A| e^{-\kappa r} \cos\left(\omega r + \delta\right) .
	\end{align}
	In this case, the behaviour in the limit is a harmonic wave shaped by a decaying exponential, which means that the asymptotic behaviour of the radial distribution function is oscillatory.
	\item The closest pole to the origin is a single real pole $s_1=-\gamma$. In that case,
	\begin{equation}\label{eq:poles_s}
		\lim_{r \rightarrow \infty} g(r) = 1+e^{s_1r} \text{Res}\left[\hat{G}(s_1)\right] ,
	\end{equation}
where $\text{Res}\left[\hat{G}(s_1)\right] = A$, which allows us to write (\ref{eq:poles_s}) as
\begin{equation}\label{eq:monotonic}
	\lim_{r \rightarrow \infty} g(r) =1+Ae^{-\gamma r} .
\end{equation}
In this second case, the behaviour of the radial distribution function for very long distances is monotonic.
\end{itemize}
\begin{figure}[htbp]
	\centering
	\includegraphics[height=7.3cm]{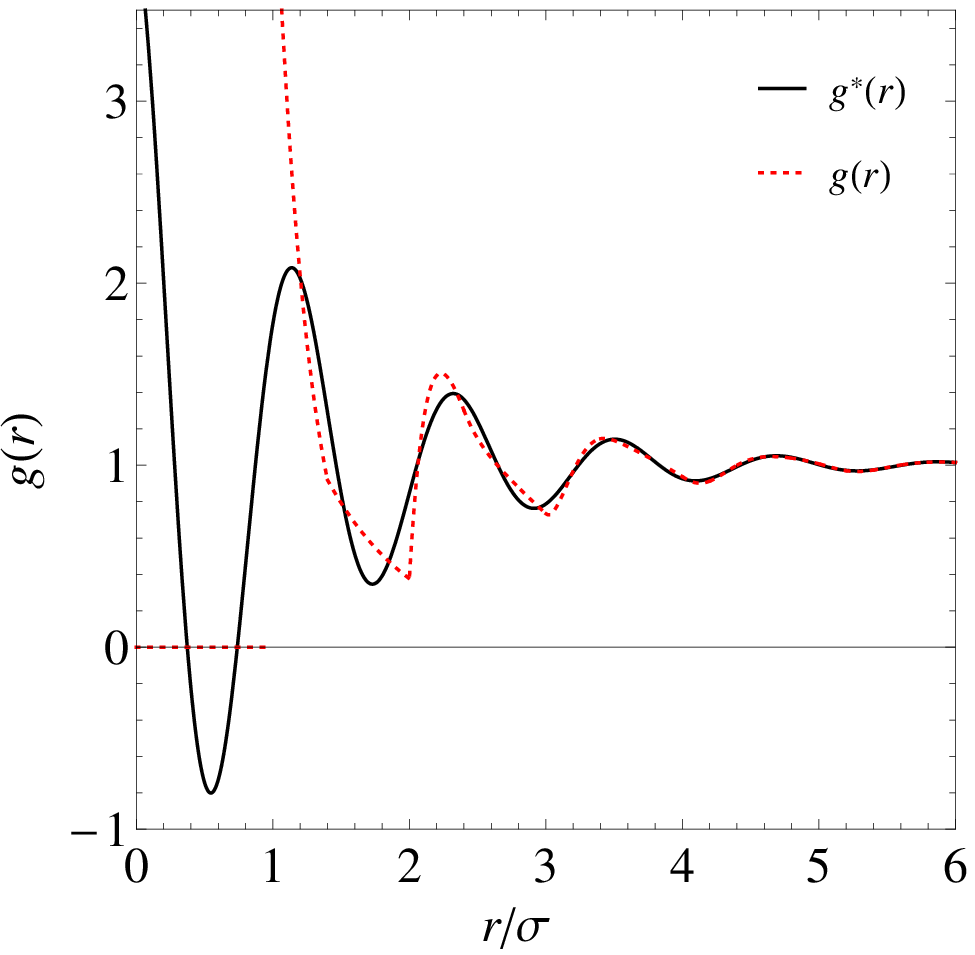}
	\caption{Plot of the radial distribution function $g(r)$ and its asymptotic approximation $g^*(r)$ at $n^*=0.6$ and a reduced temperature of $T^*=1$.} \label{fig:gr_asympt1}
\end{figure}

\begin{figure}[htbp]
	\centering
	\includegraphics[height=7.3cm]{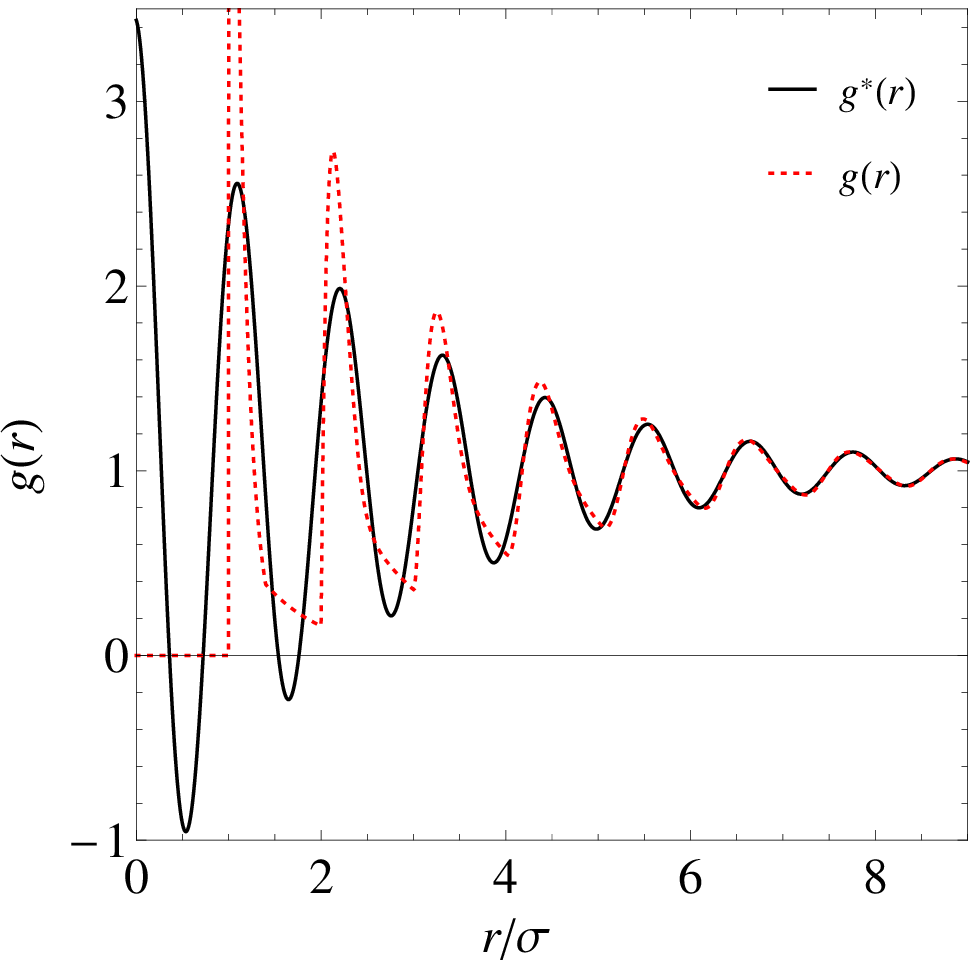}
	\caption{Plot of the radial distribution function $g(r)$ and its asymptotic approximation $g^*(r)$ at $n^*=0.6$ and a reduced temperature of $T^*=0.4$.} \label{fig:gr_asympt2}
\end{figure}

Figures~\ref{fig:gr_asympt1} and~\ref{fig:gr_asympt2} show how the radial distribution function and its asymptotic form behave: for higher temperatures, both functions are practically the same for relatively low values of $r$, while when the temperature is lower, a good agreement between both function is not achieved until we move further away from the origin. Nevertheless, and although theoretically this agreement should not be achieved until we reach very high values or $r$, these results allow us to see that this is not necessary since both functions coincide even for fairly low values of $r$.

\section{Fisher--Widom line}

In 1969, M. E. Fisher and B. Widom \cite{FW} defined a locus in the pressure-temperature plane of every attractive potential such that on one side of this locus the decay of the radial distribution function is monotonic (the closest pole to the origin is real) and on the other one it is oscillatory (the closest pole to the origin is a pair of complex conjugates). In other words, for every temperature of the system it is possible to find a transition pressure below which the decay of the radial distribution function is monotonic instead of oscillatory. Equivalently, for every temperature, there will be a pressure at which the closest pole to the origin stops being the complex one and starts being the real one.
\begin{figure}[htpb]
	\centering
	\includegraphics[height=7.3cm]{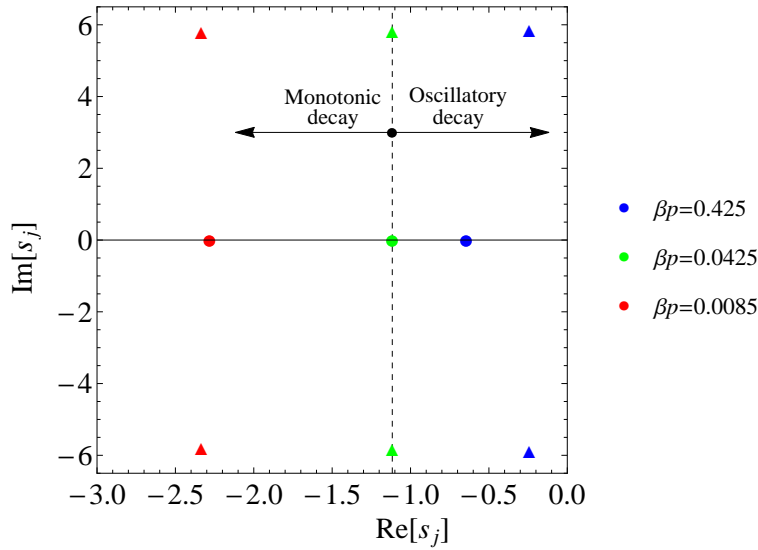}
	\caption{Plot of the behaviour of the poles of $\hat{G}(s)$ for the monotonic and oscillatory decay for a reduced temperature of $T^*=0.2$ for three different values of pressure. Complex poles are shown as triangles and real poles as circles.} \label{fig:transition_pressure}
\end{figure}
Figure~\ref{fig:transition_pressure} shows the behaviour of the poles of $\hat{G}(s)$ for different values of pressure at $T^*=0.2$: for relatively high pressures, the system presents an oscillatory decay, while if we lower the pressure enough, the position of the poles is exchanged and the system starts having a monotonic decay.

The values of the transition pressure for each temperature of the system make a curve in the pressure-temperature plane called the Fisher--Widom line. Mathematically, the conditions the poles and the transition pressure have to fulfil are the following:
\begin{gather}\label{eq:fw}
\begin{cases}
\text{Re}\left[\hat{\Omega}(-\kappa+\beta p \pm \imath \omega)\right] = \hat{\Omega}(\beta p), \\
\text{Im}\left[\hat{\Omega}(-\kappa+\beta p \pm\imath \omega)\right] = 0, \\
\hat{\Omega}(-\kappa-\beta p)= \hat{\Omega}(\beta p),
\end{cases}
\end{gather}
where $-\kappa \pm \imath \omega$ and $-\kappa$ are the complex and real pole located on the same vertical line and $\beta p$ is the value of the transition pressure.
Equation (\ref{eq:fw}) can be solved numerically but the transition pressure can only be found if the interaction potential has an attractive part. If the interaction potential is purely repulsive, then $g(r)$ decays as a damped oscillatory function for every pair of pressure-temperature values of the system.

\begin{figure}[htbp]
	\centering
	\includegraphics[scale=0.85]{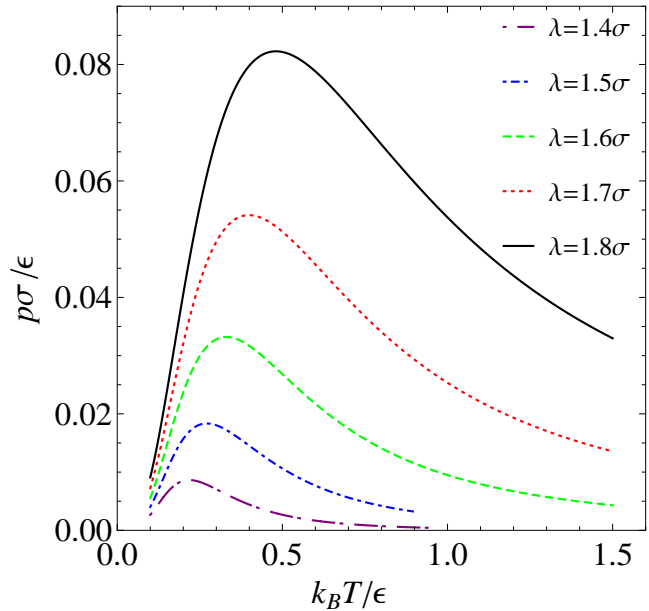}
	\caption{Fisher--Widom line for the triangle-well potential for different values of the parameter $\lambda$. The asymptotic decay of $g(r)$ is monotonic below each curve and oscillatory above it.} \label{fig:fw_line}
\end{figure}

Figure~\ref{fig:fw_line} shows the Fisher--Widom line for the triangle-well potential for different values of the parameter $\lambda$. It clearly shows that the critical pressure (the pressure above which it is not possible to have a monotonic decay, regardless of the temperature) gets bigger as the range of the potential increases. This behaviour of the critical pressure is explained by the fact that the monotonic decay of the radial distribution function is caused by the attractive nature of the potential which means that, as the range of the triangle-well potential gets larger, the monotonic behaviour of the system can be found at larger pressures.

\chapter{Ramp potential}\label{chap:ramp}

The ramp potential can be defined mathematically as
\begin{equation}\label{eq:RP}
\phi (r) =
\begin{cases}
\infty & \text{if $r\le \sigma$}, \\
\displaystyle{\frac{(r - \lambda) \epsilon}{\sigma - \lambda}} & \text{if $\sigma < r < \lambda$}, \\
0 & \text{if $r \ge \lambda$.}
\end{cases}
\end{equation}

Its graphical representation can be found in Fig.~\ref{ramp}. It easy to note that the expression for the ramp potential is formally equal to the triangle-well one [see~(\ref{eq:TW})] if we make the change $\epsilon \rightarrow -\epsilon$. This means that expressions for $\hat{\Omega}(s)$,  $\hat{\Omega}'(s)$, $\hat{\Upsilon}(s)$ and $g(r)$ are the same as in (\ref{eq:Omega}), (\ref{eq:Omega'}), (\ref{eq:Upsilon}) and (\ref{eq:rdf}), respectively, except that, in this case, the parameters $X$ and $a$ are defined as follows [see (\ref{eq:Xa})]:

\begin{equation}\label{aX-rp}
X \equiv e^{-\beta \epsilon}, \quad a \equiv \frac{\beta \epsilon}{\lambda - 1} .
\end{equation}

Apart from the formal change $\epsilon\to -\epsilon$, the ramp potential is physically very different from the triangle-well one. While the latter has an atarctive tail, the former is purely repulsive.

\section{Thermodynamic quantities}

\subsubsection*{Compressibility factor}
\begin{figure}[htpb]
	\centering
	\includegraphics[scale=0.85]{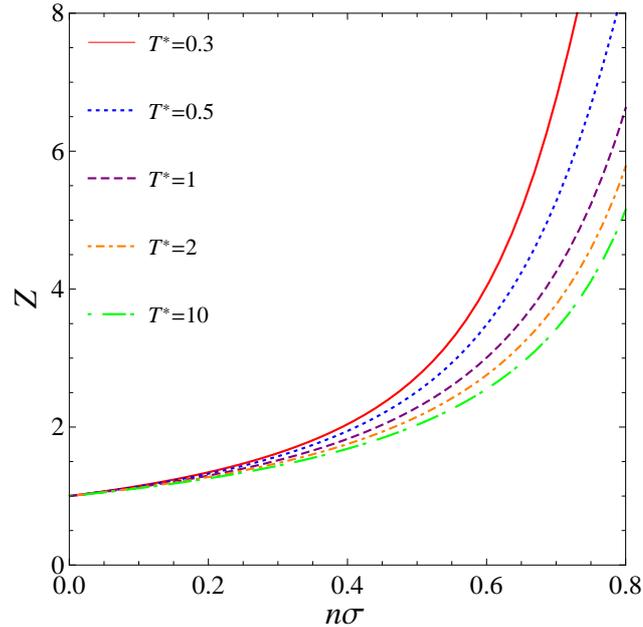}
	\caption{Density dependency of the compressibility factor $Z=\beta p/n$ of the one-dimensional ramp model for several temperatures at $\lambda=1.4\sigma$.}
	\label{fig:RP-Z}
\end{figure}

The mathematical expression for the compressibility factor in the ramp potential model is identical to (\ref{eq:density-tw}) with $a$ and $X$ defined in (\ref{aX-rp}). It is plotted in Fig.~\ref{fig:RP-Z} for some representative cases. As expected, for very low densities, the compressibility factor approaches $1$ since all systems behave like an ideal gas in the limit of low densities.

The main difference with respect to the triangle-well potential in Fig.~\ref{fig:well_z} is that, since the ramp potential is a purely repulsive model, the value of $Z$ is higher when we lower the temperature and decreases otherwise.
\subsubsection*{Excess internal energy}

The excess internal energy can be expressed as in (\ref{eq:energy-tw}) with the specific values of $a$ and $X$ for the ramp potential.

Again, because of the repulsive nature of the ramp potential, the excess internal energy is always positive instead of negative (as occurred in Fig~\ref{fig:well_energy}). Also, as expected, the excess internal energy is higher when temperature and density are also higher

\begin{figure}[htpb]
	\centering
	\includegraphics[scale=0.85]{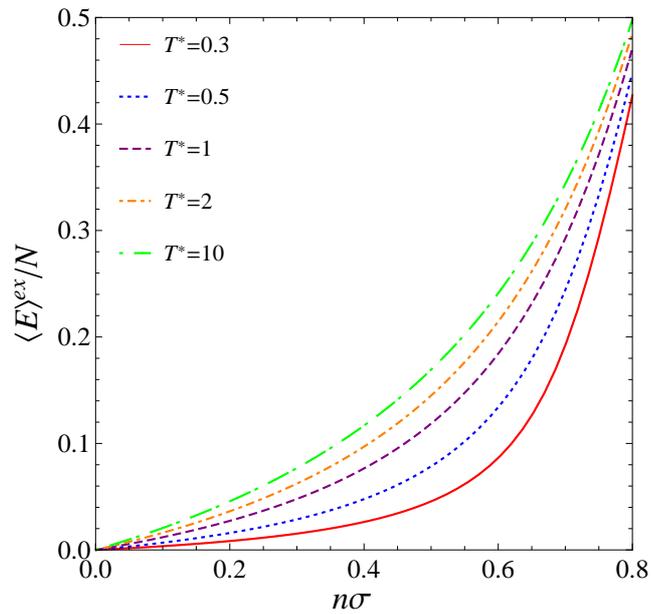}
	\caption{Density dependency of the excess internal energy of the one-dimensional ramp model for several temperatures at $\lambda=1.4\sigma$.}
	\label{fig:RP-energy}
\end{figure}

\section{Radial distribution function}

\begin{figure}[htpb]
	\centering
	\includegraphics[scale=0.85]{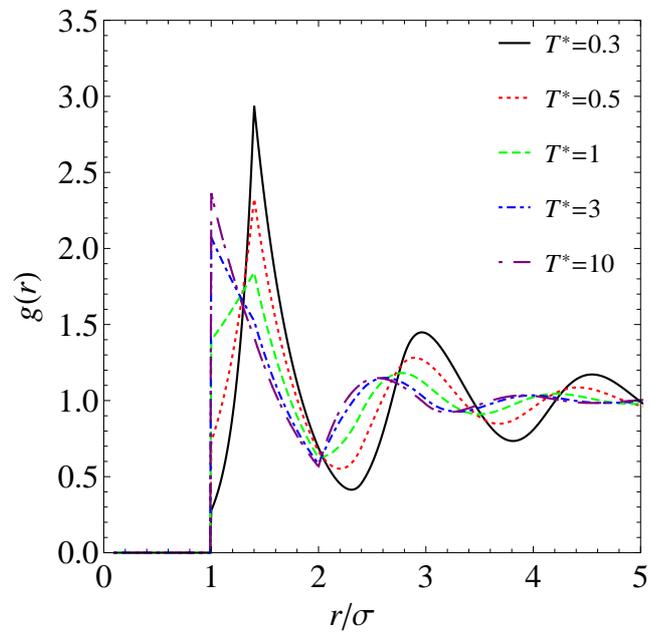}
	\caption{Plot of the radial distribution function of the one-dimensional ramp potential fluid for several temperatures at $n^*=0.6$ and $\lambda = 1.4\sigma$.}
	\label{fig:rdft-rp}
\end{figure}

\begin{figure}[htpb]
	\centering
	\includegraphics[scale=0.85]{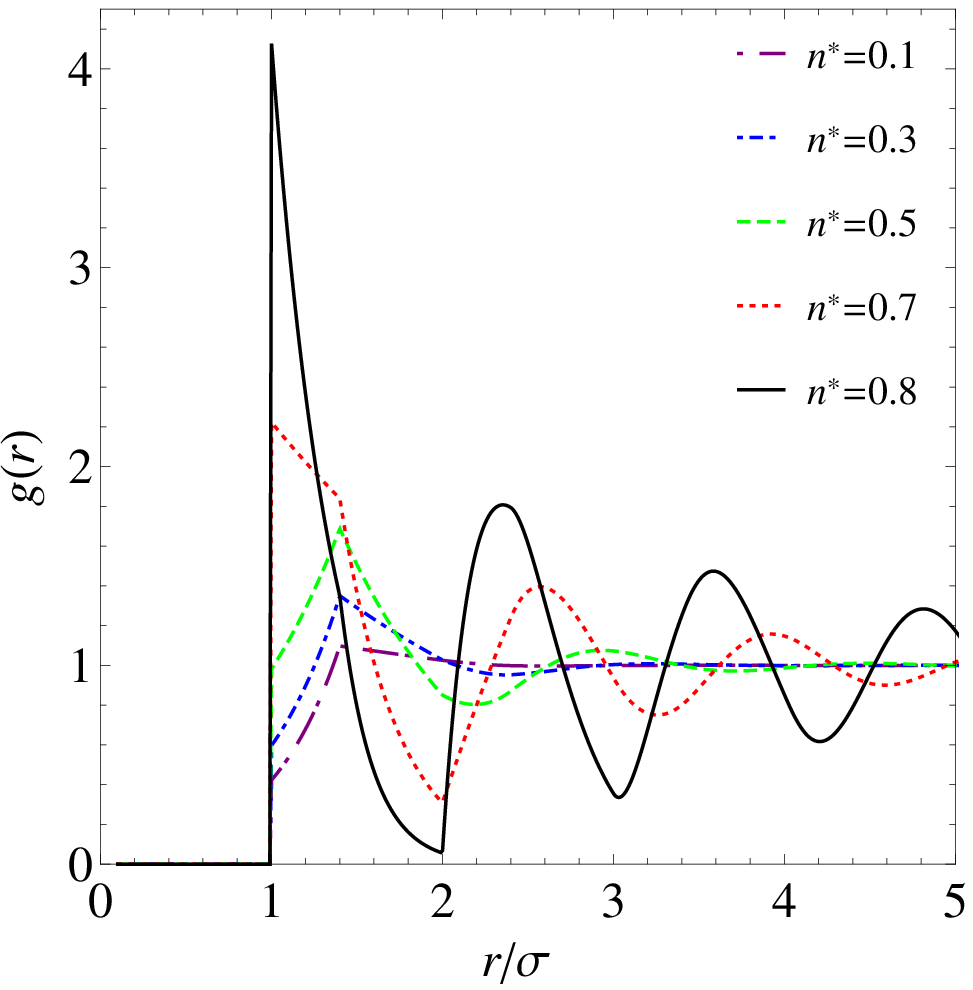}
	\caption{Plot of the radial distribution function of the one-dimensional ramp fluid for several densities at $T^*=1$ and $\lambda = 1.4\sigma$.}
	\label{fig:rdfn-rp}
\end{figure}

Figures~\ref{fig:rdft-rp} and~\ref{fig:rdfn-rp} show the radial distribution function for a ramp potential system for different temperatures and densities, respectively.

In Fig.~\ref{fig:rdft-rp} it is easy to see how, as expected, the radial distribution function goes to unity faster when the temperature of the system is higher because the ordered structure of the particles in the liquid vanishes for high temperatures. Also, the position of the particles closer to the one in the origin moves further away when the temperature is lower due to the repulsive nature of the potential. For higher temperatures, the particles pack more closely but of course never closer to the origin than a distance $r=\sigma$.

In Fig.~\ref{fig:rdfn-rp} it is possible to see how differently the system behaves for high and low densities. When density is low, the system behaves almost like an ideal gas, the only indication of the existence of a short-range order being noticeable in the region $\sigma<r<\lambda$. For higher densities, there is a well structured short-range order and the system is more closely packed, with the a sharp peak located at $r=\sigma$.

\section{Structure factor}

\begin{figure}[htpb]
	\centering
	\includegraphics[scale=0.85]{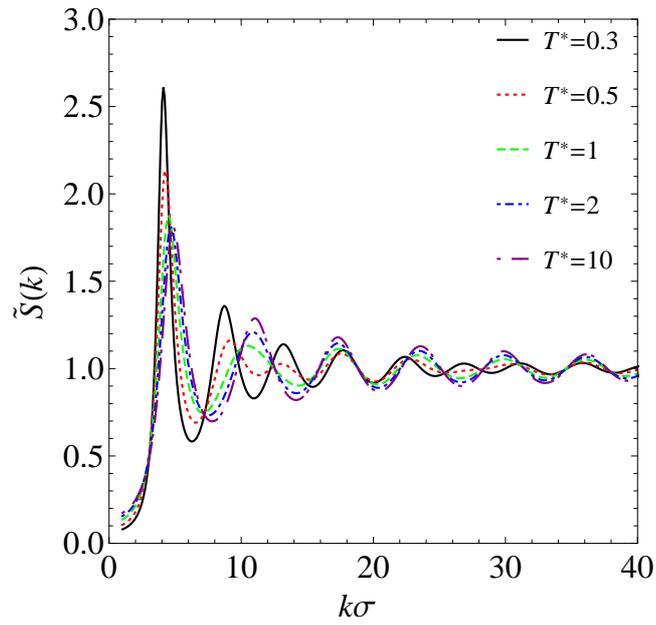}
	\caption{Plot of the structure factor of the one-dimensional ramp potential for several temperatures at $n^*=0.6$ and $\lambda = 1.4 \sigma$.}
	\label{fig:sk_T-rp}
\end{figure}

\begin{figure}[htpb]
	\centering
	\includegraphics[scale=0.85]{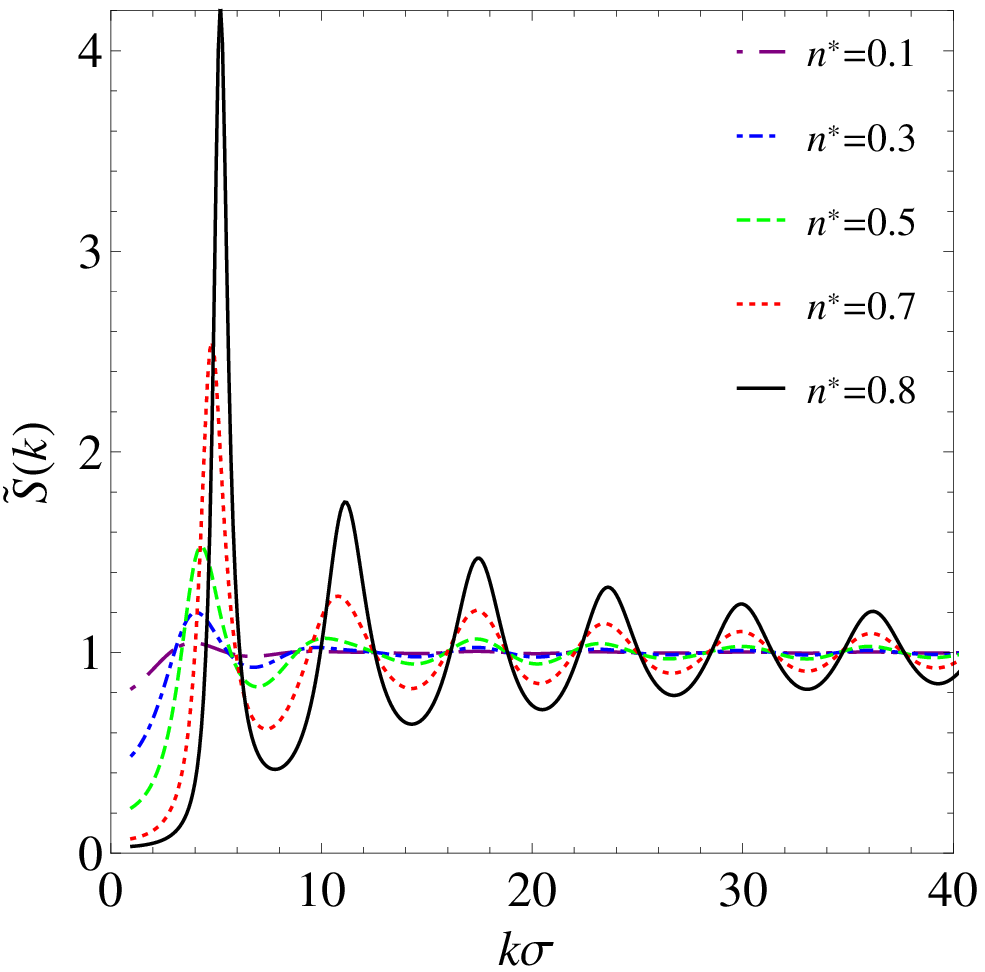}
	\caption{Plot of the structure factor of the one-dimensional ramp potential for several densities at $T^*=1$ and $\lambda = 1.4 \sigma$.}
	\label{fig:sk_n-rp}
\end{figure}

The structure factor for the ramp potential for different temperatures and densities can be seen in Figs.~\ref{fig:sk_T-rp} and~\ref{fig:sk_n-rp}, respectively. In Fig.~\ref{fig:sk_T-rp} we can see that for low temperatures the first peak is sharper but the structure factor rapidly decreases to values close to $1$. On the other hand, for high temperatures the first peak is smaller but $\tilde{S}(k)$ does not tend to $1$ so rapidly. This behaviour differs from the one found in the triangle-well potential (see Fig.~\ref{fig:sk_n}), where the value of $\tilde{S}(k)$ for high temperatures was always underneath the one for low temperatures.

The behaviour for different densities is shown in Fig.~\ref{fig:sk_n-rp} where, as expected, the structure factor is very close to unity for low densities.

\section{Direct correlation function}

\begin{figure}[htpb]
	\centering
	\includegraphics[scale=0.85]{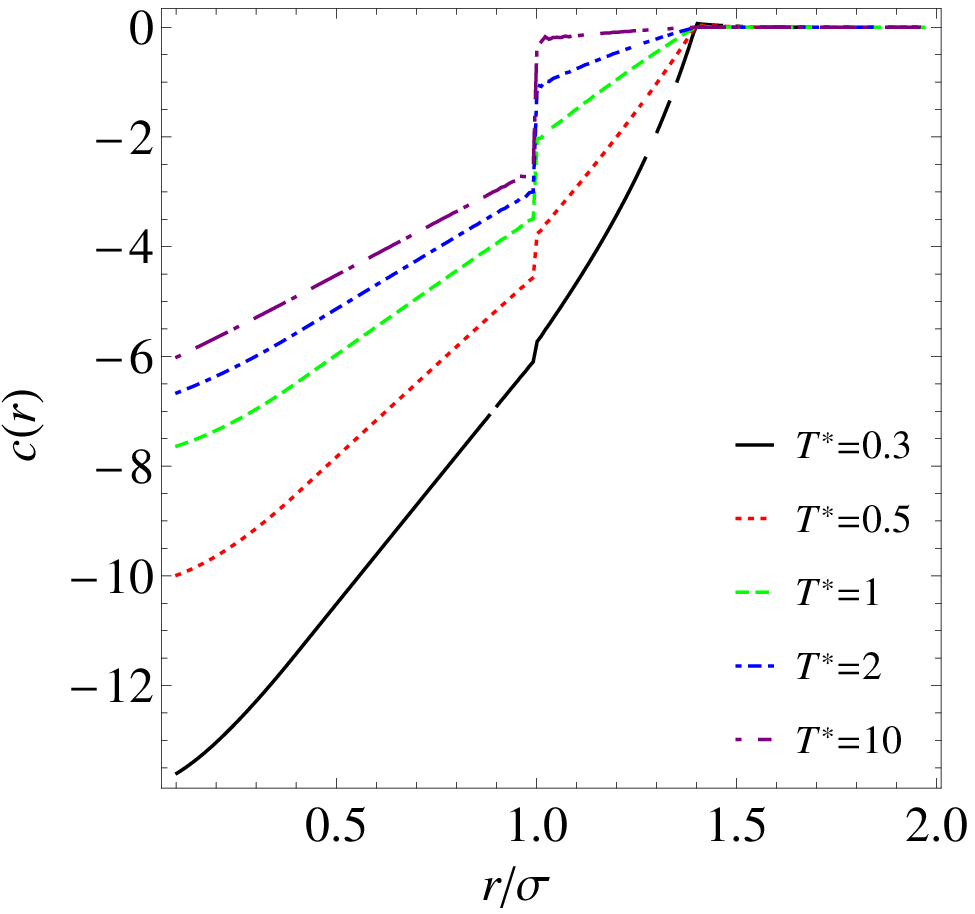}
	\caption{Plot of the direct correlation function of the one-dimensional ramp potential fluid for several temperatures at $n^*=0.6$ and $\lambda = 1.4 \sigma$.}
	\label{fig:dcfT-rp}
\end{figure}

\begin{figure}[htpb]
	\centering
	\includegraphics[scale=0.85]{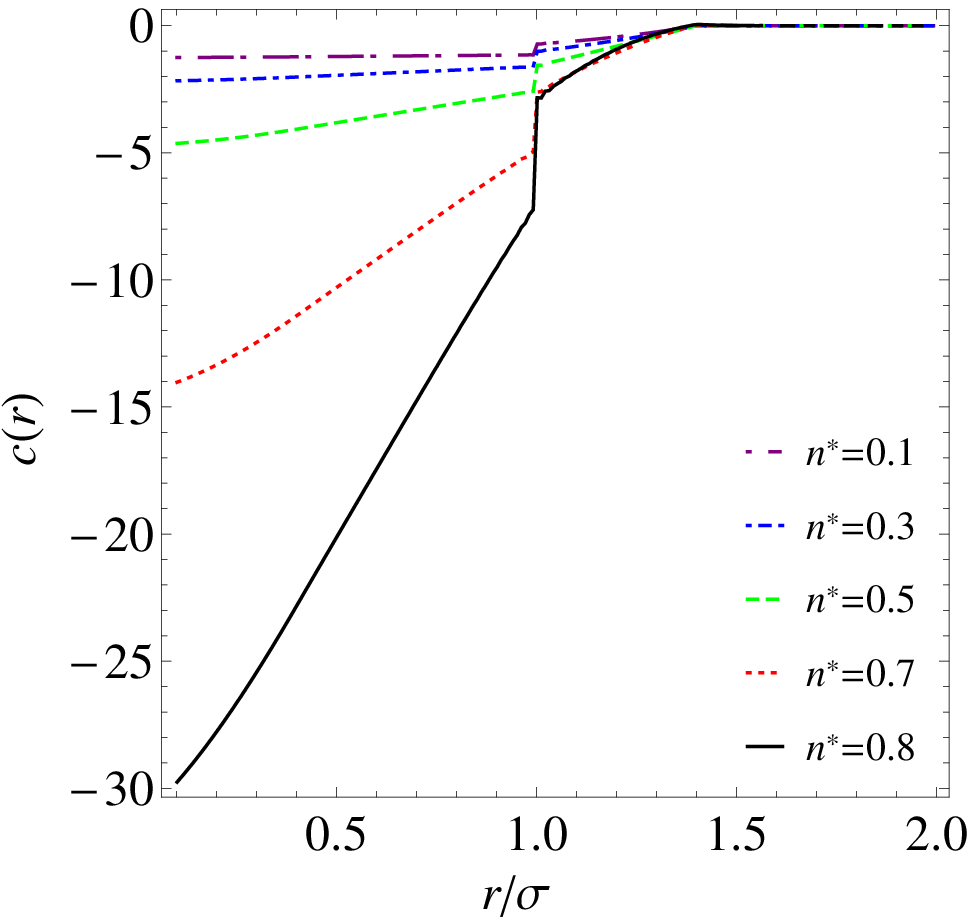}
	\caption{Plot of the direct correlation function of the one-dimensional ramp potential fluid for several densities at $T^*=1$ and $\lambda = 1.4 \sigma$.}
	\label{fig:dcfn-rp}
\end{figure}

The direct correlation function for the ramp potential is shown in Figs.~\ref{fig:dcfT-rp} and~\ref{fig:dcfn-rp}. This function presents a discontinuity at $r=\sigma$, takes in general negative values in the region $r < \sigma$ and is almost zero for $r > \sigma$. The function, however, may take positive values in the region close to $r=\sigma$ and present a small peak there that is more visible for lower temperatures than for higher ones.


\chapter{Other interaction potentials}\label{chap:last}

Varying the parameters $\lambda$ and $\sigma$ in the triangle-well potential studied in chapter~\ref{chap:tw} in the appropriate way, it is possible to obtain the radial distribution function (and all the other properties) for several different potentials. In this chapter, we will approach two different potentials: the sticky-hard-rod and the hard-rod ones.

\section{Sticky-hard-rod potential}

In 1968, Rodney J. Baxter proposed the so-called sticky-hard-sphere potential \cite{Ben-Naim}. This potential is characterized by a hard-core of diameter $\sigma$ plus an infinitely deep and infinitely narrow attractive well (see Fig.~\ref{mayerSHS}), in such a way that the second virial coefficient is finite. In the sticky-hard-sphere limit, the Mayer function becomes:
\begin{equation}\label{mayerSHS}
	f_{\mathrm{SHS}}(r) = - \Theta(\sigma -r) + \frac{\tau^{-1}}{d 2^{d-1}}\sigma \delta (r-\sigma) ,
\end{equation}
where $\tau^{-1}$ measures the ``stickiness'' of the interaction and $d$ is the dimensionality of the system.

\begin{figure}[htpb]
	\centering
	\includegraphics[scale=0.75]{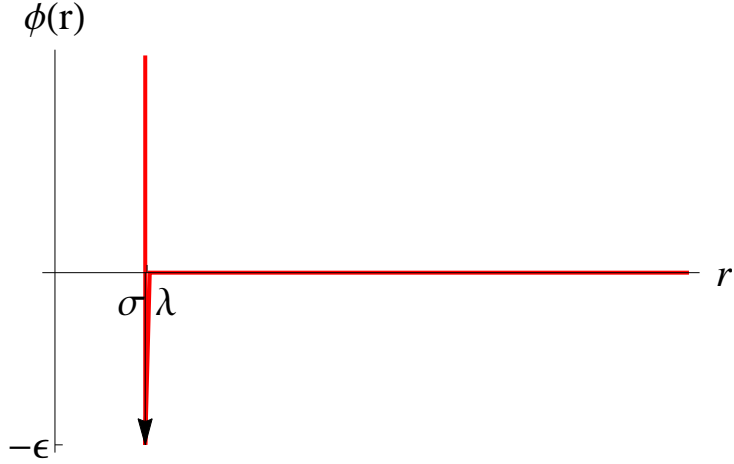}
	\caption{Sticky-hard-rod potential.}
	\label{fig:shr}
\end{figure}

In order to transform the triangle-well potential into the sticky-hard-rod one, it is necessary to make $\lambda \rightarrow \sigma$ and $\epsilon \rightarrow \infty$ but these two limits must be coupled in some way. In order to obtain the relationship between temperature and the parameter $\tau$ (at a small but finite $\lambda-\sigma$), we are going to impose the condition that the second virial coefficients for the two potentials must be equal.
Let us recall that the virial coefficients are the coefficients of the series expansion of the pressure in powers of the density \cite{Andres, Balescu, Hill, Barrio, Helfand}.

\subsubsection*{Virial coefficients for the sticky-hard-rod potential}
We consider now the sticky-hard-rod potential, which is the one-dimensional version of the sticky-hard-sphere fluid, for which the Mayer function (\ref{mayerSHS}) can be expressed as
\begin{equation}
	e^{-\beta\phi(r)}-1 = -\Theta(\sigma - r)+\tau^{-1}\sigma \delta(r-\sigma),
\end{equation}
which, taking $\sigma=1$, yields the following results:
\begin{equation}\label{eq:omega-SHS}
	\hat{\Omega}_{\mathrm{SHR}}(s) = \left(\tau^{-1} + \frac{1}{s}\right) e^{- s}
\end{equation}
and
\begin{equation}\label{eq:density-SHS}
	n=\frac{\beta p (\beta p+\tau )}{\beta p^2+\beta p \tau +\tau } .
\end{equation}

Taking into account (\ref{eq:density-SHS}), it is possible to write $\beta p$ as a series expansion in powers of $n$:
\begin{equation}
\beta p = n + \frac{\tau -1}{\tau}n^2 + \frac{\tau^2 - 2\tau +2}{\tau^2}n^3 + O(n^4) ,
\end{equation}
so the second virial coefficient for the sticky-hard-rod fluid is
\begin{equation}\label{eq:virial-SHR}
	B^{(2)}_{\mathrm{SHR}} = \frac{\tau -1}{\tau} .
\end{equation}

\subsubsection*{Virial coefficients for the triangle-well potential}

Taking into account the expression (\ref{eq:density-tw}) for the density of the triangle-well liquid as a function of pressure, it is possible to expand $\beta p$ in powers of $n$, yielding the result

\begin{equation}
	\beta p = n + \left( -T^* + e^{1/T^*}T^* + \lambda + T^* \lambda - e^{1/T^*}T^*\lambda\right)n^2 + O(n^3) ,
\end{equation}
so the second virial coefficient of the triangle-well fluid is
\begin{equation}\label{eq:virial-TW}
 B_{\mathrm{TW}}^{(2)} = -T^*(1-e^{1/T^{*}})+[1+T^{*}(1-e^{1/T^*})]\lambda .
\end{equation}

Equating~(\ref{eq:virial-SHR}) and~(\ref{eq:virial-TW}) we obtain, at a given value of $\lambda$, the relationship between temperature and the stickiness parameter:
\begin{equation}
	\tau = \frac{1}{\left(e^{1/T^*} T^*-T^*-1\right) (\lambda-1)}
\end{equation}

\begin{figure}[htpb]
	\centering
	\includegraphics[scale=0.9]{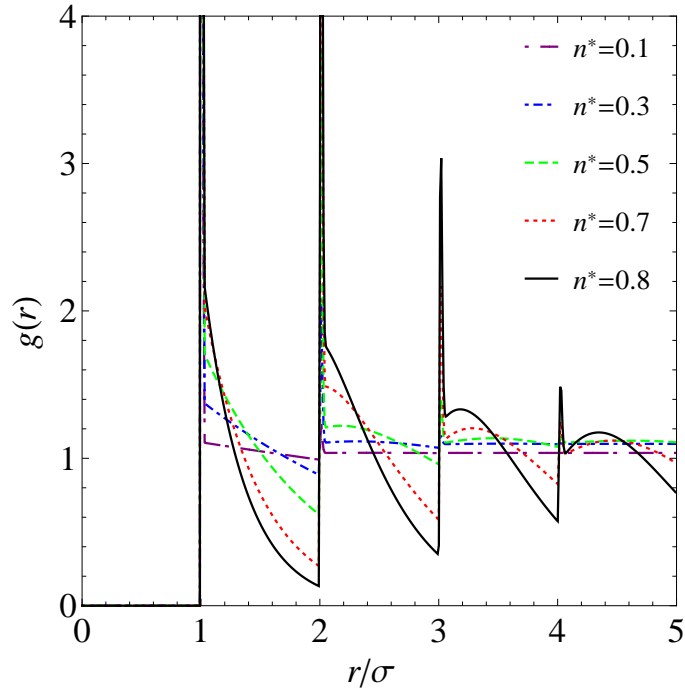}
	\caption{Radial distribution function of the one-dimensional sticky-hard-rod fluid for several densities at $\tau =5$ and $\lambda = 1.03\sigma$.}
	\label{fig:gr_shr}
\end{figure}

The radial distribution function for the sticky-hard-rod potential is represented in Fig.~\ref{fig:gr_shr}. The radial distribution function has several peaks at the values $r=\sigma, 2\sigma, 3\sigma,...$, that become smaller when we move further away from the origin.

\section{Hard-rod potential}
\begin{figure}[htpb]
	\centering
	\includegraphics[scale=0.7]{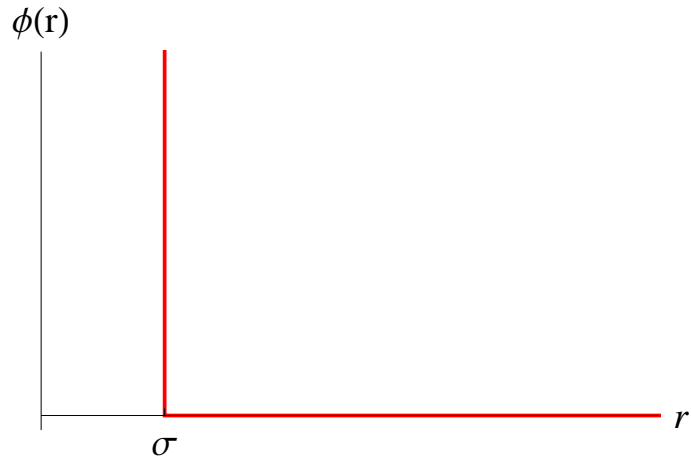}
	\caption{Hard-rod potential.}
	\label{fig:hr}
\end{figure}

\begin{figure}[htpb]
	\centering
	\includegraphics[scale=0.9]{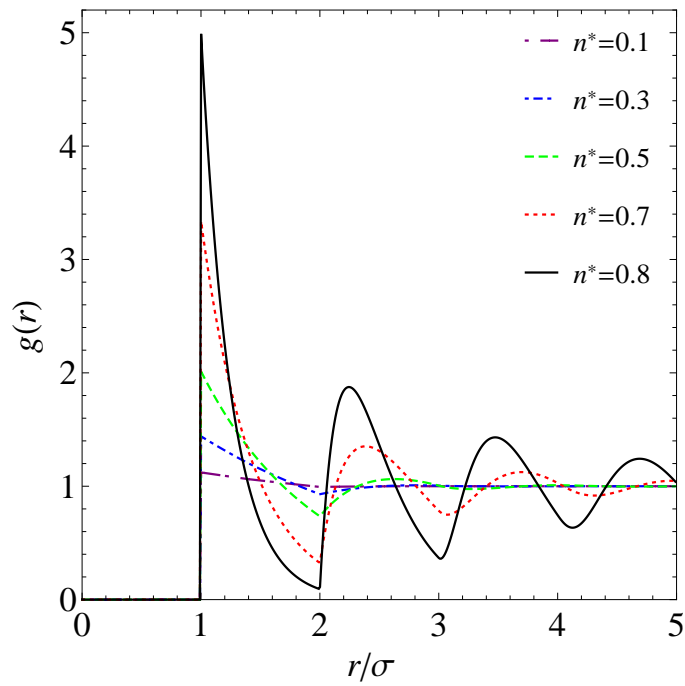}
	\caption{Radial distribution function of the one-dimensional hard-rod fluid for several densities.}
	\label{fig:hrtw}
\end{figure}

The hard-rod potential is the simplest potential in one-dimensional systems and represents impenetrable particles of diameter $\sigma$ (see Fig.~\ref{fig:hr}). It is possible to represent this potential through either the triangle-well potential or the ramp potential. In both cases, the hard-rod potential is achieved by taking the limit $\beta \epsilon \rightarrow 0$. That is, by raising the temperature of the system, relative to the energy scale $\epsilon$.
In Fig.~\ref{fig:hrtw} it is possible to observe the characteristics of the hard-rod potential, which are the same as for a system with a very high temperature.

\chapter*{Conclusions}
\markboth{Conclusions}{}

After the study of the properties of the one-dimensional triangle-well and ramp potentials and the analysis of the results in order to understand the behaviour of these classical liquids, the main conclusions that can be learnt from this project are the following ones:

\begin{enumerate}
	\item For very low densities, all the properties of the system approach those of the ideal gas, regardless of the chosen interaction potential.
	\item All liquids present a spatial short-range order that disappears when we move further away from the origin. Particles near the reference one get more ordered as we lower the temperature or increase the density.
	\item The Percus--Yevick equation is, in general, a better approximation to the analytical direct correlation function from the triangle-well than the hypernetted-chain, although both of them work well if the density of the system is sufficiently low.
	\item A good agreement between the radial distribution function of the triangle-well and its asymptotic behaviour (far away from the origin) is achieved for relatively low distances from the origin.
	\item The critical pressure for the Fisher--Widom line in the triangle-well gets higher as the range of the potential increases, which is in accordance with the fact that the bigger the attractive nature of a potential, the higher the critical pressure of the Fisher--Widom line.
	
	\item It is also possible to study accurately certain simple interaction potentials by means of the results of other ones just by making the appropriate changes in the parameters describing the potential.
\end{enumerate}
\addcontentsline{toc}{chapter}{Conclusions}


\cleardoublepage
\addcontentsline{toc}{chapter}{Bibliography}
\bibliographystyle{unsrt}
\bibliography{biblio2}

\end{document}